\documentclass[10pt,journal,doublecolumn]{IEEEtran}
\usepackage{amsmath,amssymb,threeparttable,placeins,enumerate,caption}
\usepackage{mathtools,relsize}
\usepackage[pdftex]{graphicx}
\usepackage{epstopdf}
\usepackage{color}
\usepackage[usenames]{xcolor}
\usepackage{amsfonts}
\usepackage{latexsym}
\usepackage{subfigure,multirow,makecell,stfloats}
\usepackage{subfloat}
\usepackage[none]{hyphenat}
\usepackage{makecell}

\usepackage{algorithm}
\usepackage[noend]{algpseudocode}

\def\qi#1 {\fbox {\footnote {\ }}\ \footnotetext { From Qi: {\color{red}#1}}}

\newtheorem{theorem}{{{\textit{Theorem}}}}

\newtheorem{lemma}{{{\textit{Lemma}}}}
\newtheorem{corollary}{{{{\textit{Corollary}}}}}

\newtheorem{definition}{{{\textit{Definition}}}}

\newtheorem{remark}{{{\textit{Remark}}}}

\newtheorem{example}{{{\textit{Example}}}}

\newtheorem{Construction}{Construction}
\begin{document}

\title{Low Ambiguity Zone: Theoretical Bounds and Doppler-Resilient Sequence Design in Integrated Sensing and Communication Systems}

\author{Zhifan~Ye,~
        Zhengchun~Zhou,~\IEEEmembership{Member,~IEEE,}
         Pingzhi~Fan,~\IEEEmembership{Fellow,~IEEE,}
        Zilong~Liu,~\IEEEmembership{Senior~Member,~IEEE,}
         Xianfu~Lei,~\IEEEmembership{Member,~IEEE,}
         and~Xiaohu~Tang,~\IEEEmembership{Senior~Member,~IEEE}

       \thanks{Manuscript received August 11, 2021; revised December 8, 2021;
accepted January 14, 2022. The work of Zhifan Ye and Zhengchun Zhou was supported by the NSFC
project No.62071397 and No. 62131016, and in part by  projects of central government to guide local scientific and technological development under Grant No.2021ZYD0001.. The work of Pingzhi Fan was supported by the NSFC
project No.62020106001 and the 111 project No.111-2-14. This article (as an invited paper) was presented in part at the 2021 IEEE/CIC International Conference on Communications in China (ICCC), Xiamen, China, in July 2021 [DOI: 10.1109/ICCC52777.2021.9580208]. (Corresponding
author: Zhengchun Zhou.)}

\thanks{Z. Ye is with the School of Mathematics, Southwest Jiaotong University, Chengdu, 611756, China. E-mail: yzffjnu@163.com.}
\thanks{Z. Zhou is with the Key Lab of Information Coding and Wireless Comms., and also with the School of Mathematics, Southwest Jiaotong University, Chengdu, 611756, China. E-mail: zzc@swjtu.edu.cn.}
\thanks{Z. Liu is with the School of Computer Science and Electronics Engineering, University of Essex, Colchester CO4 3SQ, U.K.. E-mail: zilong.liu@essex.ac.uk.}
\thanks{P. Fan, X. Lei and X. Tang are with the Key Lab of Information Coding and Wireless Comms., Southwest Jiaotong University, Chengdu, 611756, China. E-mail: pzfan@swjtu.edu.cn; xflei@swjtu.edu.cn; xhutang@swjtu.edu.cn.}
}
\maketitle

\date{}

\begin{abstract}
In radar sensing and communications, designing Doppler resilient sequences (DRSs) with low ambiguity function for delay over the entire signal duration and Doppler shift over the entire signal bandwidth is an extremely difficult task. However, in practice, the Doppler frequency range is normally much smaller than the bandwidth of the transmitted signal, and it is relatively easy to attain quasi-synchronization for delays far less than the entire signal duration. Motivated by this observation, we propose a new concept called low ambiguity zone (LAZ) which is a small area of the corresponding ambiguity function of interest defined by the certain Doppler frequency and delay. Such an LAZ will reduce to a zero ambiguity zone (ZAZ) if the maximum ambiguity values of interest are zero. In this paper, we derive a set of theoretical bounds on periodic LAZ/ZAZ of unimodular DRSs with and without spectral constraints, which include the existing bounds on periodic global ambiguity function as special cases. These bounds may be used as theoretical design guidelines to measure the optimality of sequences against Doppler effect. We then introduce four optimal constructions of DRSs with respect to the derived ambiguity lower bounds based on some algebraic tools such as characters over finite field and cyclic difference sets.
\end{abstract}

\begin{IEEEkeywords}
Radar Sensing and Communications, Ambiguity Function, Low Ambiguity Zone, Theoretical Bounds, Doppler Resilience, Spectral Constraints, High Mobility Communications, Optimal Sequences.
\end{IEEEkeywords}

\section{Introduction}
\subsection{Background}

In recent years, integrated sensing and communication (ISAC) systems capable of simultaneously performing sensing and communication tasks while sharing the same hardware and bandwidth resources have attracted increasing research attention \cite{Ma2020}. A typical application scenario of ISAC is for connected autonomous vehicles where multiple sensors and communication devices are jointly deployed to measure/track/exchange certain key system information such as speed, vibration, approaching target distance. These ISAC systems enjoy mutually enhanced radar and communication functions compared to a single radar or communication system. Specifically, rapid and high-resolution radar sensing result can be attained by utilizing some important information (e.g., geographical locations, carrier frequency and bandwidth, moving speeds and routes) acquired by the communication module. Also, the communication quality can be improved, for example, by adjusting the antenna directions for more accurate beamforming or by better compensating the Doppler estimated from the radar sensing result.

That said, legacy sensing and wireless communication systems have been studied as separate research entities.
A sensing system often operates in an environment which is corrupted by noise and may consist of a variety of clutters. Thus, two primary goals of any sensing system are to attain enhanced parameter estimation and target resolution. The parameter estimation of a target includes information related to its size, motion, and location \cite{Skolnik}. The ability of one or more sensors to distinguish different targets that are very close is characterized by target resolution.
On the other hand, wireless communication systems are required to provide certain minimum quality-of-services (QoSs) to meet the user demands in various applications \cite{Molisch}.
Thanks to the advancement in digital circuit technology,
most of the system functions of integrated radar sensing and communication applications can be realized using adaptive and configurable software systems while keeping the same RF front-end architecture \cite{Tavik2005}.
From the waveform design point of view, there are four main types of ISAC schemes \cite{Ma2020,Hassanien2019}: 1) coexistence schemes, which utilize independent waveforms for each functionality (sensing or communication); 2) communications waveform-based approaches, where communication signals such as orthogonal frequency-division multiplexing (OFDM) and orthogonal time-frequency-space (OTFS) are used for sensing; 3) radar waveform-based schemes, in which digital messages are embedded into radar waveforms; and 4) joint waveform design approaches, which achieve the joint sensing-communication system by deploying dedicated dual-function waveforms. This paper is mainly concerned with the radar waveform-based ISAC scheme in which the waveforms exhibiting strong resilience to Doppler are needed.

\begin{figure*}[!htbp]
\setlength{\belowcaptionskip}{-0.5cm}   
  \centering
  \centering
  {\includegraphics[height=6cm,width=15cm]{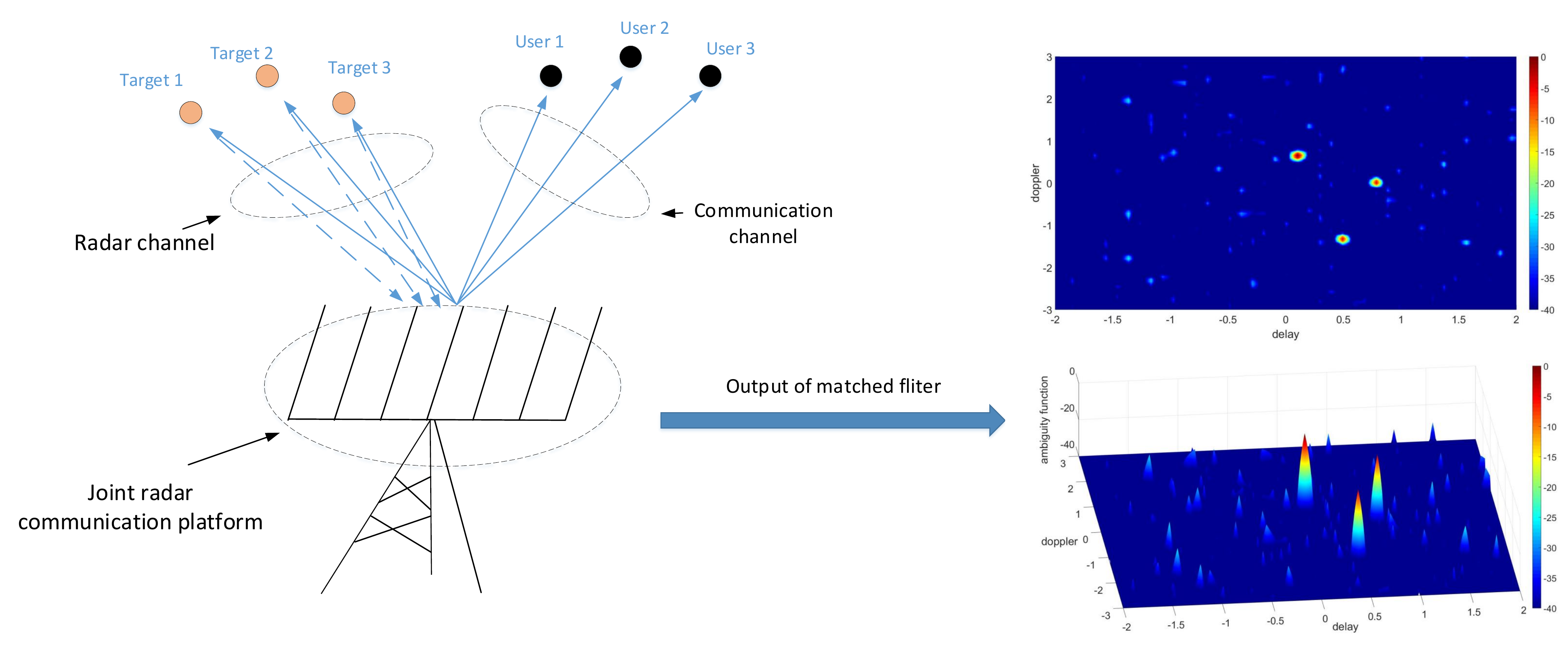}}
    \caption{A diagram of an ISAC system.}
\end{figure*}
By generating appropriate waveforms, one may change radar waveforms from pulse to pulse, and employ the waveform itself as a means of embedding communication symbols \cite{Hassanien2019,Blunt2010,Ted2018}, e.g., code shift-keying (CSK) where each waveform corresponds to a code representing a communication symbol.
Implementing such a CSK scheme requires careful waveform design to ensure that their auto- and cross- ambiguity function (AF) should be as small as possible, so as to attain high-resolution radar sensing performance.
From the communications perspective, one needs to deal with the notorious Doppler effect incurred by high mobility especially in high frequency bands \cite{Fan2016}.
Thus, it is of strong interest to study the fundamental limits and systematic constructions of sequences which are robust to Doppler. Such sequences having low AF values are called Doppler-resilient sequences (DRSs) \cite{Pezeshki}.
Throughout this work, sometimes, they may also be referred to as the ISAC waveforms.

\subsection{Motivation}

As illustrated in Figure 1, both sensing and communication systems are integrated in a single hardware platform, with the same waveform and the same transmitter. Consider a joint sensing-communication platform equipped with a number of transmit antennas arranged as a uniform linear array. The sensing receiver employs an array of receive antennas with an arbitrary linear configuration, while the communication receivers are assumed to be located in the direction known to the transmitter. The information data rate is mainly determined by the radar pulse repetition frequency (PRF), whether the system uses a phased-array or multiple-input, multiple-output (MIMO) configuration, and the permissible incremental changes in radar waveform structure and bandwidth. Note that an AF represents the time response of a filter matched to a signal when the signal is received with a delay and a Doppler shift \cite{Levanon2004}. A good AF may be represented by a spike that peaks at the origin and extremely small values in other delay-Doppler areas. A waveform with such AF property gives rise to excellent parameter estimation and favourable resolution between neighboring targets in the region. However, designing DRSs with such globally low ambiguity function is an extremely difficult task. In many practical applications \cite{Pezeshki,Arlery2016,Kum2018,Duggal2021}, the Doppler frequency range can be much smaller than the bandwidth of the transmitted signal, and it may not be necessary to consider the whole signal duration.
As shown in Figure 1, the three targets can be distinguished effectively if the AF has very low sidelobes in the region of interest.
Such a small region with low/zero ambiguity function values is called low/zero ambiguity zone (LAZ/ZAZ) in this paper.

The concept of LAZ/ZAZ, explicitly quantifies the local region of interest, i.e., the low/zero ambiguity zone (as a design parameter).
To derive the related theoretical bounds with respect to the quantified local zone and provide theoretical design guidance to such DRSs, it is essential to understand the the tradeoff between the sequence length, the set size, and the maximum ambiguity function value. In the literature, theoretical bounds and optimal constructions of sequences with low/zero correlation properties have been extensively investigated \cite{Helleseth1998}. By contrast, the theoretical limits on AF are much more complicated due to the two-dimensional shifts in delay-Doppler domain. To the best of our knowledge, only Ding \emph{et al.} generalized the maximum AF lower bound in \cite{Ding1}.
Moreover, very few works are known on DRSs meeting the derived bound in \cite{Ding1}. Although there are some works on sensing and communications such as \cite{Garmatyuk2008,Garmatyuk2011,Vossiek2003,Chen2021} which mainly
focus on the design, analysis and optimization of practical ISAC systems, a comprehensive study on the fundamental limits of ISAC waveforms is missing.

Besides, the design of ISAC waveforms needs to take into account of the availability and contiguousness of the spectrum. As spectrum may be underutilized most of the time, it is likely for secondary users to reuse certain spectrum holes in an opportunistic manner for improved spectrum utilization efficiency \cite{Liu}. Specifically, the idle spectrum resources can be detected by spectrum sensing and then allocated to secondary users with minimum interference to the primary ones.

\subsection{Contributions and Organization}

The main contributions of this paper are summarized as follows.
\begin{itemize}
  \item To provide more signal design freedom, i.e. more desirable sequences (waveforms) for integrated sensing and communication systems, a new concept called LAZ/ZAZ, corresponding  to a small area of interest defined by the maximum Doppler frequency and the maximum delay is proposed.
      Further, theoretical bounds on periodic LAZ of unimodular DRSs with respect to the zone size of LAZ/ZAZ, the sequence length, and the number of sequences are derived. It is shown that the obtained bounds include the existing bounds on periodic global ambiguity function as special cases. The theoretical bound on LAZ/ZAZ shows that one can improve the Doppler resistance at the expense of appropriate delay resistance, or vice versa.
  \item To investigate the impact of spectral constraints, i.e., non-contiguous spectral bands, on sequence design, theoretical bounds on periodic LAZ/ZAZ of unimodular DRSs with spectral constraints are derived as well. It is also shown that these bounds include the existing periodic global bounds as special cases. The results indicate that theoretical bounds on ZAZ are consistent under spectral constraints or non-spectral constraints.
        \item To validate the LAZ/ZAZ concepts and the above theoretical bounds, four systematic constructions of DRSs with and without spectral constraints, which are optimal with respect to the derived bounds, are presented, based on some algebraic tools such as characters over finite field and cyclic difference sets.
\end{itemize}

The remainder of this paper is organized as follows. A new concept called low/zero ambiguity zone (LAZ and ZAZ) is introduced, as shown in Section II. In Section III, we introduce some famous correlation bound of traditional sequence and some known DRSs. In Section IV, we aim to derive four types of lower bounds on periodic LAZ/ZAZ of unimodular DRSs with and without spectral constraints. These bounds are tight in the sense that they can be achieved with equality by some optimal DRSs, which are presented in Section V and Section VI. Some examples and potential applications of LAZ/ZAZ DRSs are provided in Section VII.

\section{Ambiguity function and low/zero ambiguity zone}

AF, defined as a two dimensional delay-Doppler correlation function of the transmitting signals, plays a central role in waveform design and performance evaluation. The mainlobe width, sidelobe level, and ambiguity peaks of an AF have direct impact on the range resolution, sidelobe interference, and ambiguity characteristics.

There are two types of AF: aperiodic AF and periodic AF. In the literature, an AF delay may range the whole signal duration $T$ (or full period in the case of periodic AF), and the AF Doppler shift may span the whole signal bandwidth $1/T$. In general, the ideal aperiodic/periodic AF should exhibit a single sharp peak at the origin (which is the nominal delay and Doppler for that matched filter), and small values close to zero elsewhere (thumbtack shape). However, as the volume underneath an AF square is a constant \cite{Levanon2004}, if the AF value is lowered in certain area of the delay-Doppler plane, it may rise somewhere else.
To cope with such scenarios and to more efficiently design the desired signals, this paper proposes a new concept called LAZ, corresponding to a small area of interest defined by the maximum Doppler frequency and the maximum delay (instrumented range).

With the aid of massive MIMO, the communication rate of ISAC system can be improved greatly \cite{Hassanien2019}. Let us assume that the received signal model at receiver $n$ due to the signal transmitted from transmitter $m$ is
\begin{equation}
y_{n, m}(t)=\alpha_{n, m} s_{m}\left(t-\tau_{n, m}\right) e^{j 2 \pi f_{n, m} t}+z_{n, m}(t),
\end{equation}
where $\tau_{n, m}, f_{n, m}$ and $\alpha_{n, m}$ represent the time delay, Doppler shift and reflection coefficients, respectively, corresponding to the path between the $m$-th transmitter and the $n$-th receiver, and $z_{n, m}(t)$ denotes the noise. After passing through the matched filter bank, the core of signal processing is the calculation of AF.
For a pair of DRSs $(\mathbf{a},\mathbf{b})$ of length $N$, the discrete aperiodic cross AF is defined as follows:
\begin{equation}
\widetilde{AF}_{\mathbf{a},\mathbf{b}}(\tau, \nu)=\left\{\begin{array}{ll}
\sum\limits_{t=0}^{N-1-\tau} a(t) b^{*}(t+\tau) e^{j 2 \pi \nu t/N},\\ \quad\quad\quad\quad\quad\quad\quad\quad\quad\quad  0 \leq \tau \leq N-1;
\\
\sum\limits_{t=-\tau}^{N-1} a(t) b^{*}(t+\tau) e^{j 2 \pi \nu t/N}, \\\quad\quad\quad\quad\quad\quad\quad\quad\quad\quad  1-N \leq \tau<0;
 \vspace{0.2cm}\\
0, \quad\quad\quad\quad\quad\quad\quad\quad\quad  |\tau| \geq N.
\end{array}\right.
\end{equation}
Specially, the aperiodic cross AF shall become periodic cross AF when the summation variable $t=0$ to $N-1$ (modulo $N$), that is,
\begin{equation}\label{AF}
AF_{\mathbf{a},\mathbf{b}}(\tau, \nu)=\sum\limits_{t=0}^{N-1} a(t) b^{*}(t+\tau) e^{j 2 \pi \nu t/N},
\end{equation}
where $\tau,\nu$ are called time- and Doppler- shifts, respectively, $|\tau|, |\nu|\in \mathbb{Z}_{N}$, $j=\sqrt{-1}$ and the summation $t+\tau$ is modulo $N$. If $\mathbf{a}=\mathbf{b}$, we call it auto-ambiguity function denoted by $AF_{\mathbf{a}}(\tau, \nu)$. This article will focus on investigating periodic AF with the aid of certain algebraic tools. The maximum ambiguity magnitude of DRS family $\mathcal{S}$ is defined as
$
\theta_{\max}=\max \{\theta_{A},\theta_{C}\},
$
where the maximal auto-ambiguity magnitude
\small{
\begin{equation}
\theta_{A}=\max\Bigl \{\left|AF_{\mathbf{a}}(\tau,\nu)\right|: \mathbf{a} \in \mathcal{S}, (0,0)\neq(|\tau|,|\nu|)\in\mathbb{Z}_{N}\times\mathbb{Z}_{N}\Bigl \},
\end{equation}}
and the maximal cross-ambiguity magnitude
\small{
\begin{equation}
  \theta_{C}=\max \Bigl\{\left|AF_{\mathbf{a},\mathbf{b}}(\tau,\nu)\right|: \mathbf{a}\neq\mathbf{b} \in \mathcal{S}, 0\leq |\tau| <N, 0\leq |\nu| <N\Bigl \}.
\end{equation}
}
The ambiguity magnitude of DRS family $\mathcal{S}$ of size $M$ over a region $\Pi\subseteq (-N,N)\times(-N,N)$ can be defined as
\begin{equation}
\begin{split}
  F_{\Pi}(\mathcal{S})=\max \Bigl \{\left|AF_{\mathbf{a}, \mathbf{b}}(\tau,\nu)\right|&:\mathbf{a}, \mathbf{b} \in \mathcal{S} ~\text {and}~ (\tau,\nu)\in \Pi,\\
  &(0,0)\neq (\tau,\nu)\in \Pi  \text { if } \mathbf{a}=\mathbf{b}\Bigl \}.
\end{split}
\end{equation}
\begin{figure*}
\setlength{\belowcaptionskip}{-0.5cm}   
  \centering
\begin{minipage}[t]{0.5\textwidth}
\centering
\includegraphics[width=8cm]{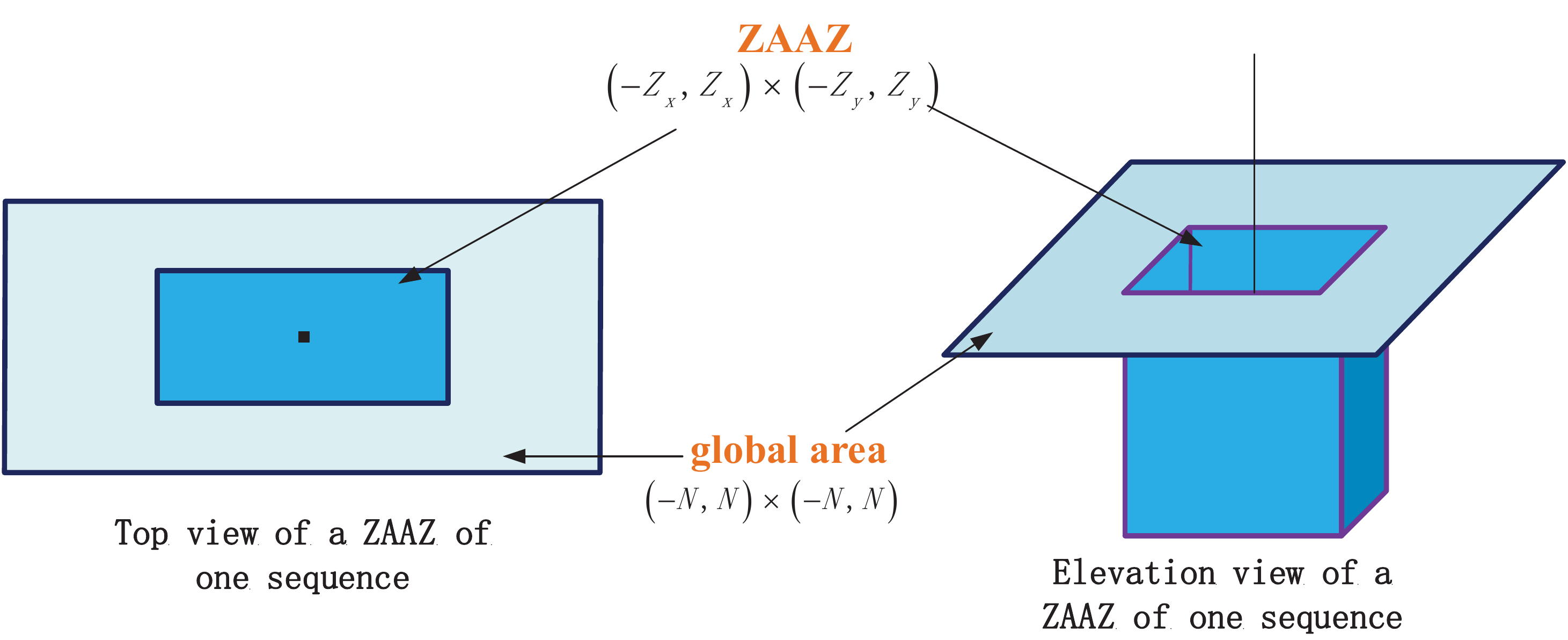}
\caption{Illustration of a ZAAZ of one sequence.}
\label{a}
\end{minipage}%
\begin{minipage}[t]{0.5\textwidth}
\centering
\includegraphics[width=8cm]{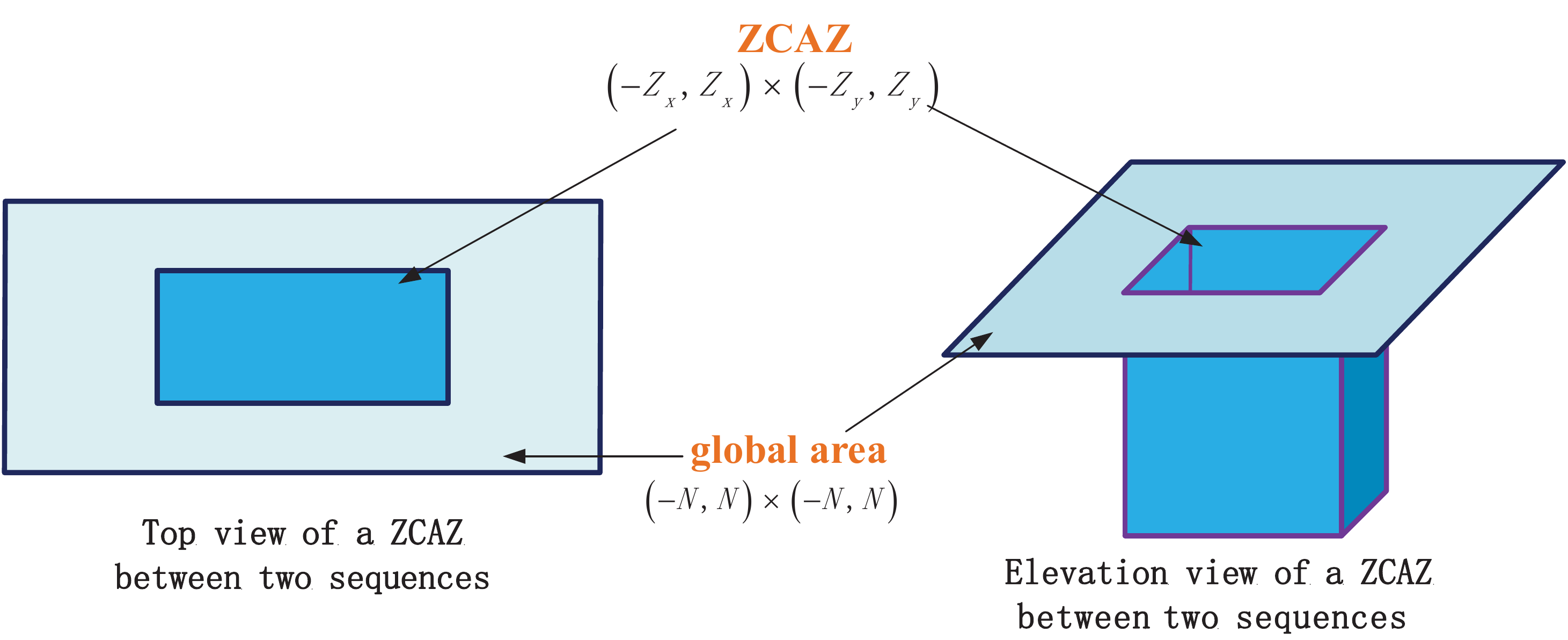}
\caption{Illustration of a ZCAZ between two sequences.}
\label{b}
\end{minipage}
\end{figure*}
Such a DRS set with maximum ambiguity magnitude $\theta_{\max}=F_{\Pi}(\mathcal{S})$ over region $\Pi$ is denoted by $(N,M,\theta_{\max},|\Pi|)$-$\mathcal{S}$, where $|\Pi|$ is the area of $\Pi$. In particular, we sometimes drop off $\Pi$ and denote the DRS set by $(N,M,\theta_{\max})$-$\mathcal{S}$ if we consider global ambiguity function with $\Pi=(-N,N)\times(-N,N)$. It is noted that conventional correlation function is a special case of the ambiguity function when $\nu=0$. In wireless communications and radar sensing, low ambiguity is expected at any given delay and Doppler shift in order to detect and identify targets and achieve information transmission between devices. However, designing such codes with low ambiguity function for delay over whole signal duration and Doppler shift over whole signal bandwidth is an extremely difficult task.
Obviously finding DRSs with LAZ characteristics is a relatively feasible task. In fact, one can even design DRSs having periodic zero ambiguity zone. Now we give  a formal definition of periodic low/zero ambiguity zone.

\begin{definition}
Let $\mathcal{S}$ be a set consisting of $M$ distinct sequences of period $N$, for a small nonnegative real number $\theta$, the periodic low ambiguity zone with maximum ambiguity magnitude $\theta$ for sequence set $\mathcal{S}$ is defined as
\begin{equation}\label{eqn-LAZ}
\begin{split}
  \Pi_{\theta}&(\mathcal{S})= \Bigl \{\Pi:F_{\Pi}(\mathcal{S})\leq \theta\Bigl \}.
\end{split}
\end{equation}
In practice, the LAZ is required as large as possible. With this goal in mind, define
\begin{equation}\label{LAZ}
\begin{split}
  \Pi_{\max}= \Bigl \{\Pi\in \Pi_{\theta}&(\mathcal{S}):|\Pi|=\max\{|A|:A\in \Pi_{\theta}(\mathcal{S})\}\Bigl \},
\end{split}
\end{equation}
such a sequence set is denoted by $(N,M,\theta,\Pi_{\max})$ DRS set.
Specially, if $\theta=0$, that corresponds to the periodic zero ambiguity zone. Similarly, aperiodic LAZ and  $(N,M,\widetilde{\theta},\Pi_{\max})$ aperiodic DRS set can be defined.
\end{definition}

An ideal ambiguity function may be represented by a spike that peaks at the origin and takes zero everywhere. Such an ambiguity function provides perfect resolution between neighboring targets regardless of how close they are to
each other. An ideal ambiguity function does not exist due to the bound to be derived in this paper. Yet, similar to the existing sequences with zero correlation zone (ZCZ) \cite{Fan1999}, it is possible to construct a set of sequences which possess zero auto-ambiguity zone (ZAAZ) and zero cross-ambiguity zone (ZCAZ), as illustrated in Fig. 2 and Fig. 3.

\section{A Review of Celebrated Correlation Bounds and DRSs}
A conventional correlation is a one-dimensional function of time-shifts. Let us define the (periodic) cross-correlation function between sequences $\mathbf{a}=[a(0), a(1),\cdots, a(N-1)]$ and $\mathbf{b}=[b(0),b(1),\cdots,b(N-1)]$ with period $N$ as
\begin{equation}
R_{\mathbf{a}, \mathbf{b}}(\tau)=\sum\limits_{t=0}^{N-1} a(t) b^{*}(t+\tau),
\end{equation}
where $\mathbf{a}$ and $\mathbf{b}$ are complex unit sequences, and $t+\tau$ is calculated over modulo $N$. If $\mathbf{a}=\mathbf{b}$, the cross-correlation function of $\mathbf{a}$ and $\mathbf{b}$ reduces to the auto-correlation function of $\mathbf{a}$, denoted by $R_{\mathbf{a}}(\tau)$. Let $\mathcal{S}$ be a set consisting of $M$ distinct complex unit sequences of period $N$. Then the maximal auto-correlation magnitude of $\mathcal{S}$ is defined as
\begin{equation}
\lambda_{A}=\max \{\left|R_{\mathbf{a}}(\tau)\right|: \mathbf{a} \in \mathcal{S}, 0< |\tau| <N\}.
\end{equation}
Similarly, the maximal cross-correlation magnitude of $\mathcal{S}$ is defined as
\begin{equation}
\lambda_{C}=\max \{\left|R_{\mathbf{a}, \mathbf{b}}(\tau)\right|: \mathbf{a} \neq \mathbf{b} \in \mathcal{S}, 0\leq |\tau| <N\}.
\end{equation}
Moreover, the maximal correlation magnitude of $\mathcal{S}$ is defined as
\begin{equation}
\lambda_{\max}=\max \{\lambda_{A},\lambda_{C}\}.
\end{equation}
The set $\mathcal{S}$ is called a complex unimodular $(N,M,\lambda_{\max})$ sequence family of period $N$ with the family size $M$ and the maximum correlation $\lambda_{\max}$. In \cite{Welch}, Welch developed a lower bound on $\lambda_{\max}$ given by
\begin{equation}
\lambda_{\max}\geq N\sqrt{\frac{M-1}{NM-1}}.
\end{equation}
In 1979, Sarwate \cite{Sarwate} established a trade-off of $\lambda_{A}$ and $\lambda_{C}$, i.e.,
\begin{equation}
\frac{N-1}{N(M-1)} \cdot\left(\frac{\lambda_{A}^{2}}{N}\right)+\left(\frac{\lambda_{C}^{2}}{N}\right) \geq 1.
\end{equation}
These bounds are important guidelines for conventional sequence design in which Doppler is not a major concern. A number of optimal conventional sequence sets with respect to these bounds have been reported in the literature (see \cite{Popovic1,Alltop,Zhou,Gol} and references therein). Based on the Welch bound, Tang and Fan 
derived the following set size upper bound of zero correlation zone sequences which are useful in quasi-synchronous code-division multiple-access (QS-CDMA) systems:
$
MZ_{cz}\leq N,
$
where
\begin{equation}
\begin{split}
Z_{cz} =\mbox{max}\Bigl \{ T: R_{{\bf a}, {\bf b}}(\tau) = 0 ~~\mbox{for}~ &0 <\vert \tau \vert < T~\mbox{or}~  \tau = 0 \\
& \mbox{and}~  {\bf a} \neq  {\bf b},  \forall ~{\bf a}, {\bf b} \in\mathcal{S}\Bigl \}.
\end{split}
\end{equation}
So far, many families of ZCZ sequence sets have been reported in the literature \cite{Tangbound,Chen,Fan1,Zhou2}.

It is noted that traditional sequence design generally assumes the availability of a contiguous spectral band, meaning that the sequence energy can be allocated to all the carriers of such a spectral band. Such a design paradigm may not be able to carry on anymore due to the increasingly congested spectrum. Recently, significant research attention has been paid on sequence design over non-contiguous spectrum bands. Liu \emph{et al.} developed in \cite{Liu} a series of periodic and aperiodic-correlation lower bounds for spectral constrained sequences (SCSs) by carrying out convex optimization in the frequency domain. Specifically, consider $\mathbf{C}_{i}=[C_{i}(0),C_{i}(1),\ldots,C_{i}(N-1)]$ which is the frequency domain dual of an SCS with energy $N$, i.e.,
$
\left|C_{i}(f)\right|^{2}=0,~\text{if}~f\in \Omega,
$
where $\Omega\subseteq\{0,1,\ldots,N-1 \}$ is called the spectral constraint set which consists of all the forbidden carrier indices. For a frequency-domain DRS dual set $\mathcal{C}=\{\mathbf{C}_{0},\mathbf{C}_1,\cdots,\mathbf{C}_{M-1}\}$, we assume $|\Omega|=N-L$, where $L$ denotes the number of all the admissible carriers over which active power transmission are allowed. The maximum periodic correlation magnitude of an $(N,M,\lambda_{\max})$ SCS family $\mathcal{S}$ satisfies the following lower bound\footnote{This is essentially (44) of \cite{Liu}.}
\begin{equation}
\lambda_{\max}\geq N\sqrt{\frac{N(M-1)+N-L}{L(NM-1)}}.
\end{equation}
The above bound reduces to the well-known Welch bound when the number of forbidden carriers is set to zero, i.e., $L=N$.
In 2013, Ding \emph{et al.} generalized the Welch bound for DRS sets \cite{Ding1}. Formally, for any $(N,M,\theta_{\max})$ DRS set $\mathcal{S}$, one has
\begin{eqnarray}\label{welch-AF}
\theta_{\max}\geq N\sqrt{\frac{NM-1}{N^{2}M-1}}.
\end{eqnarray}

The study of DRSs has been attracting increasing research
attention in recent years. Ding \textit{et al.} proposed a class of
sequences which asymptotically meets the Welch bound for
DRSs in \cite{Ding1}. With the aid of additive character and multiplicative
character over finite field, Wang and Gong constructed
several families of polyphase sequences having low ambiguity
amplitudes \cite{Gong,Wang2011,Wang}. Schmidt provided a direct proof for the
Wang-Gong construction by the Weil bound of hybrid character
sums \cite{K.U.}. Using the theory of finite-unit norm tight frames,
Benedetto and Donatelli computed the ambiguity amplitudes
of the Frank-Zadoff-Chu sequences in \cite{Ben}.
In addition, there are many researches to shape local or global ambiguity function by optimization algorithm in the literature (\cite{Stein,Cui,Jing} and references therein).

Whilst each of the above DRSs is designed with the availability
of a contiguous spectral band, we are also concerned
with DRSs which are subject to certain spectral hole constraints
(SHSs). In modern communication and radar systems
operating in congested white space, an SHS often occurs
when certain non-contiguous carriers are not allowed for active
power allocation. A sequence satisfying an SHS is called
a spectrally-constrained sequence (SCS). In \cite{Tsai}, correlation
lower bounds and constructions of SCSs are studied in the
context of non-contiguous orthogonal frequency division multiplexing
(OFDM) systems. These bounds are extended and
generalized by Liu \emph{et al.} to single- and multi- channel SCSs in
\cite{Liu}. Moreover, two classes of optimal unimodular SCSs whose
spectral hole positions are cyclic difference sets are proposed.
Recently, a new optimal SCS set with comb-like spectral
hole positions is developed in \cite{LYB}. Popovic constructed a
family of SCSs with flexible parameters based on modulatable
constant-amplitude zero-autocorrelation (CAZAC) sequences
\cite{Popovic}. Despite these extensive research attempts, SCSs which
exhibit Doppler resilience have not been reported, to the best
of our knowledge.

\section{Bounds on Periodic LAZ/ZAZ of Unimodular DRS sets}
\subsection{Lower Bound of DRSs with LAZ/ZAZ}
Unimodular DRSs, in which each sequence is polyphase consisting of complex-valued elements with absolute value of one, are highly desirable in many communication systems for maximum power transmission efficiency. Before the context of the derivations of the main theorem, we present below a lemma to show some properties of ambiguity function for unimodular DRS family.
Proofs of all lemmas are presented in the appendix.
\begin{lemma}
For any unimodular DRS $\mathbf{u}$, we have
\begin{equation}\label{polyphase}
\left|AF_{\mathbf{u}}(0, \nu)\right|=0,~\text{for}~\nu\neq 0.
\end{equation}
\end{lemma}

Now we present our main theorem and a simple derivation by forming a ``fat" matrix which consists of all the possible time- and Doppler- shifted versions of sequences. Such a derivation sheds some light for obtaining the other bounds in this paper.

\begin{theorem}\label{th-local}\textbf{(Main Theorem)}
For any $(N,M,\theta_{\max},|\Pi|)$ unimodular DRS set $\mathcal{S}$, where $\Pi=(-Z_{x},Z_{x})\times(-Z_{y},Z_{y})$, we have
\begin{equation}\label{LAZ-bound}
\theta_{\max}\geq \frac{N}{\sqrt{Z_{y}}}\sqrt{\frac{MZ_{x}Z_{y}/N-1}{MZ_{x}-1}}.
\end{equation}
Specially, if $\theta_{\max}=0$, it reduces to
\begin{equation}
MZ_{x}Z_{y}\leq N.
\end{equation}
Therefore, the area of ZAZ $\Pi_{max}$ subjects to
\begin{equation}
\left|\Pi_{max}\right|\leq \frac{4N}{M}.
\end{equation}
\end{theorem}
\begin{IEEEproof}
Let $\mathcal{S}=\{\mathbf{u}_{i}:1\leq i \leq M\}$ with $\mathbf{u}_{i}=[u_{i}(0),u_{i}(1),$ $\ldots,u_{i}(N-1)]$. Define the following matrix from the obtained DRSs as
$
\mathbf{U}_{(Z_{x},Z_{y})}=\left [\mathbf{U}_{1}^{(Z_{x},Z_{y})},\mathbf{U}_{2}^{(Z_{x},Z_{y})},\ldots,\mathbf{U}_{M}^{(Z_{x},Z_{y})}\right ],
$
where
$\mathbf{U}_{j}^{(Z_{x},Z_{y})}=\left [\mathbf{U}_{j}^{(Z_{x})}(0),\mathbf{U}_{j}^{(Z_{x})}(1),\ldots,\mathbf{U}_{j}^{(Z_{x})}(Z_{y}-1)\right ],$ and
\begin{equation}\label{ap1}
\setlength{\arraycolsep}{0.1pt}
\begin{array}{lc}
    \mathbf{U}_{j}^{(Z_{x})}(\nu_{i})=\vspace{0.5ex}\\
     \begin{bmatrix}
     u_{j}(0)&u_{j}(1)e^{\frac{j2\pi \nu_{i}}{N}}  & \dots &u_{j}(Z_{x}-1)e^{\frac{j2\pi \nu_{i}(Z_{x}-1)}{N}} \\
u_{j}(1)e^{\frac{j2\pi \nu_{i}}{N}} & u_{j}(2)e^{\frac{j2\pi 2\nu_{i}}{N}} & \dots &u_{j}(Z_{x})e^{\frac{j2\pi \nu_{i}Z_{x}}{N}}\\
\vdots  & \vdots  & \ddots & \vdots \\
u_{j}(N-1)e^{\frac{j2\pi \nu_{i}(N-1)}{N}} &  u_{j}(0) & \dots  & u_{j}(Z_{x}-2)e^{\frac{j2\pi \nu_{i}(Z_{x}-2)}{N}}
 \end{bmatrix}
 \end{array}
\end{equation}
Note that the following identity holds:
\begin{equation}\label{ZCZ}
\parallel \mathbf{U}_{(Z_{x},Z_{y})}^{H}\mathbf{U}_{(Z_{x},Z_{y})} \parallel_{F}^{2}=\parallel \mathbf{U}_{(Z_{x},Z_{y})}\mathbf{U}_{(Z_{x},Z_{y})}^{H} \parallel_{F}^{2},
\end{equation}
where $\parallel \cdot \parallel_{F}$ stands for the Frobenius norm, and $(\cdot)^{H} $ denotes the transpose conjugate operator.
The right-hand side term of the equation (\ref{ZCZ}) is equal to
 \begin{equation}\label{row}
 \begin{split}
   &N(MZ_{x}Z_{y})^{2}\\
   &+\sum_{l\neq m=0}^{N-1}\left|\sum_{j=1}^{M}\sum_{\nu_{i}=0}^{Z_{y}-1}
\sum_{\tau=0}^{Z_{x}-1}u_{j}(l+\tau) u^{*}_{j}(m+\tau) e^{\frac{j2\pi \nu_{i}(l-m)}{N}}\right|^{2}.
\end{split}
 \end{equation}
Hence,
 \begin{equation}\label{golrowip}
\parallel \mathbf{U}_{(Z_{x},Z_{y})}\mathbf{U}_{(Z_{x},Z_{y})}^{H} \parallel_{F}^{2}\geq N(MZ_{x}Z_{y})^{2}.
\end{equation}
On the other hand, the left-hand side term of (\ref{ZCZ}) can be expanded as follows:
\begin{equation}\label{bound1}
\begin{split}
&\sum_{j=1}^{M}\sum_{\tau=1-Z_{x}}^{Z_{x}-1}\sum_{\nu=1-Z_{y}}^{Z_{y}-1}\big(Z_{x}-|\tau|\big)\big(Z_{y}-|\nu|\big)\left|AF_{\mathbf{u}_{j}}(\tau,\nu)\right|^{2}\\
&+\sum_{j,k=1\atop j\neq k}^{M}\sum_{\tau=1-Z_{x}}^{Z_{x}-1}\sum_{\nu=1-Z_{y}}^{Z_{y}-1}\big(Z_{x}-|\tau|\big)\big(Z_{y}-|\nu|\big)\left|AF_{\mathbf{u}_{j},\mathbf{u}_{k}}(\tau,\nu)\right|^{2}.
\end{split}
\end{equation}
Based on Lemma 1, we have
\begin{equation}
\begin{split}
\sum_{j=1}^{M}\sum_{\nu=1-Z_{y}}^{Z_{y}-1}Z_{x}&\big(Z_{y}-|\nu|\big)\left|AF_{\mathbf{u}_{j}}(0,\nu)\right|^{2}\\
&=MZ_{x}Z_{y}\left|AF_{\mathbf{u}_{j}}(0,0)\right|^{2}=MZ_{x}Z_{y}N^{2}.
\end{split}
\end{equation}
Therefore, we obtain
\begin{equation}\label{lasteq}
MZ_{x}Z_{y}N^{2}+MZ_{x}Z_{y}(MZ_{x}Z_{y}-Z_{y})\theta_{\max}^{2}\geq N(MZ_{x}Z_{y})^{2}.
\end{equation}
The result then follows.
\end{IEEEproof}

As a special case, some known bounds can be derived from Theorem 1, we give the following corollary to illustrate.

\begin{corollary}
With the same notations as before.
\begin{itemize}
\item When $Z_{x}=Z_{y}=N$, a lower bound of global ambiguity is given as
\begin{eqnarray}\label{eqn-bound1}
\theta_{\max}\geq \sqrt{N}.
\end{eqnarray}
In addition, we can get the Sarwate bound of DRS set as follows
\begin{equation}
\frac{(N-1)\theta_{A}^{2}}{(MN-1)N}+\frac{(M-1)\theta_{C}^{2}}{MN-1} \geq 1.
\end{equation}

\item When $Z_{y}=1$, (\ref{LAZ-bound}) reduces to the Tang-Fan-Matsufuji bound for ZCZ sequences in \cite{Tangbound}.

\end{itemize}
\end{corollary}

\begin{remark}
The global ambiguity function lower bound in (\ref{welch-AF}) is also a special case of \textbf{Theorem 1}, which can be derived from (\ref{ZCZ}), (\ref{bound1}), (\ref{glob}) without the action of Lemma 1. However, (\ref{eqn-bound1}) in Corollary 1 is strictly tighter than the lower bound in (\ref{welch-AF}). Such a lower bound has not been reported before, to the best of our knowledge. In the next section of this paper, we will show that this lower bound is tight in the sense that one can construct an infinite family of sequences meeting the equality of (\ref{eqn-bound1}).
\end{remark}

\subsection{Lower Bound for SCSs with LAZ/ZAZ}
We aim to derive a lower bound of SCSs with LAZ/ZAZ in this subsection. Firstly, we present the ambiguity function in frequency domain based on the symmetric discrete Fourier transform, i.e.,
\begin{equation}
\begin{split}
  A(f)=\frac{1}{\sqrt{N}}\sum\limits_{t=0}^{N-1} a(t) e^{-j 2 \pi f t/N},
 \\
  a(t)=\frac{1}{\sqrt{N}}\sum\limits_{f=0}^{N-1} A(f) e^{j 2 \pi f t/N},
\end{split}
\end{equation}
where $\mathbf{A}$ denotes the frequency domain dual sequence of $\mathbf{a}$ in (\ref{AF}). Similarly, we define $\mathbf{B}$ for the sequence $\mathbf{b}$. Therefore, we have

\begin{equation}\label{FAF}
\begin{split}
AF_{\mathbf{a},\mathbf{b}}(\tau, \nu)
=\sum\limits_{f=0}^{N-1} A(f-\nu) B^{*}(f) e^{-j 2 \pi f \tau/N},
\end{split}
\end{equation}
where $|\tau|, |\nu|\in \mathbb{Z}_{N}$. Let us consider the frequency domain dual sequence $\mathbf{C}_{j}=[C_{j}(0),\ldots,C_{j}(N-1)]$ of a SCS $\mathbf{c}_{j}$ with $L\leq N$ admissible carriers. That is,
\begin{equation}
\left|C_{j}(f)\right|^{2}=
\left\{\begin{array}{ll}
\frac{N}{L}, & f \not\in \Omega;  \vspace{1ex} \\
0, & \text { otherwise. }
\end{array}\right.
\end{equation}
From now on, by substituting $\mathbf{C}_{j}$ for $\mathbf{u}_{j}$, we will analyze ambiguity function in the frequency domain with the technique in the proof of Theorem 1. Unlike the aforementioned cases, the derivation of lower bound of SCSs with LAZ is more complex. This is because the right-hand side term of (\ref{ZCZ}) is hard to determine, and we only have
\small{
\begin{equation}\label{qo}
\begin{split}
\parallel \mathbf{U}_{(Z_{x},Z_{y})}\mathbf{U}_{(Z_{x},Z_{y})} ^{H} \parallel_{F}^{2} \geq\sum_{f=0}^{N-1}\left(\sum_{j=1}^{M} Z_{x}\sum_{\nu=0}^{Z_{y}-1}|C_{j}(\nu+f)|^{2}\right)^{2}.
\end{split}
\end{equation}
}
Based on the above analysis, we present a lower bound for SCSs with LAZ as follows.

\begin{theorem}
For any $(N,M,\theta_{\max},|\Pi|)$ SCS set $\mathcal{S}$, where $\Pi=(-Z_{x},Z_{x})\times(-Z_{y},Z_{y})$, we have
\begin{equation}
\theta^{2}_{\max}\geq\frac{N(MZ_{x}Z_{y}-N)}{MZ_{x}Z_{y}-1}.
\end{equation}
Specially, if $\theta_{\max}=0$, it reduces to
\begin{equation}
MZ_{x}Z_{y}\leq N.
\end{equation}
Therefore, the area of ZAZ $\Pi_{max}$ also subjects to
\begin{equation}
\left|\Pi_{max}\right|\leq \frac{4N}{M}.
\end{equation}
\end{theorem}
\begin{IEEEproof}
Based on Cauchy-Schwarz inequality and (\ref{qo}), we have
\begin{equation}
\begin{split}
\parallel \mathbf{U}_{(Z_{x},Z_{y})}\mathbf{U}_{(Z_{x},Z_{y})} ^{H} &\parallel_{F}^{2} \geq\\
&\frac{1}{N}\cdot N\sum_{f=0}^{N-1}\left(\sum_{j=1}^{M} Z_{x}\sum_{\nu=0}^{Z_{y}-1}|C_{j}(\nu+f)|^{2}\right)^{2}\\
&\geq \frac{1}{N}\left(\sum_{f=0}^{N-1}\sum_{j=1}^{M} Z_{x}\sum_{\nu=0}^{Z_{y}-1}|C_{j}(\nu+f)|^{2}\right)^{2}\\
&=\frac{1}{N}\left(\sum_{j=1}^{M} Z_{x}\sum_{\nu=0}^{Z_{y}-1}\sum_{f=0}^{N-1}|C_{j}(\nu+f)|^{2}\right)^{2}\\
&=N(MZ_{x}Z_{y})^{2}.
\end{split}
\end{equation}
It is similar with the proof of Theorem 1, based on (\ref{ZCZ}) and (\ref{bound1}), we have
\begin{equation}
  MZ_{x}Z_{y}N^{2}+MZ_{x}Z_{y}(MZ_{x}Z_{y}-1)\theta_{\max}^{2}\geq N(MZ_{x}Z_{y})^{2}.
\end{equation}
Hence,
\begin{equation}
\theta^{2}_{max}\geq\frac{N(MZ_{x}Z_{y}-N)}{MZ_{x}Z_{y}-1}.
\end{equation}
\end{IEEEproof}

\begin{remark}
As far as we know, \cite{Tsai} is the only known work which considered the lower bound of SCS with LCZ. Specifically, it may be seen as a special case of our derived bound in Theorem 2.  By setting $Z_{y}=1$, (\ref{qo}) becomes
\begin{equation}
\begin{split}
  \sum_{f=0}^{N-1}\left(\sum_{j=1}^{M} Z_{x}|C_{j}(f)|^{2}\right)^{2} =\frac{N^{2}(MZ_{x})^{2}}{L}.
\end{split}
\end{equation}
Then, by (\ref{ZCZ}) and (\ref{bound1}), we have
\begin{equation}
 MZ_{x}N^{2}+MZ_{x}(MZ_{x}-1)\lambda_{\max}^{2}\geq \frac{N^{2}(MZ_{x})^{2}}{L}.
\end{equation}
Hence,
\begin{equation}
\lambda^{2}_{max}\geq\frac{N^{2}(MZ_{x}-L)}{L(MZ_{x}-1)},
\end{equation}
which is the lower bound in \cite{Tsai}.
\end{remark}

When $Z_{x}=Z_{y}=N$, equations (\ref{ZCZ})-(\ref{bound1}) still hold and the right-hand side term of (\ref{ZCZ}) reduces to  $N(MN^{2})^{2}$. In this case, the left-hand side term of (\ref{ZCZ}) can be estimated by the property of correlation for SCS family in \cite{Liu} as follows
\begin{equation}
\begin{split}
\sum_{\tau=0}^{N-1}\left|AF_{\mathbf{c}_{j},\mathbf{c}_{k}}(\tau,0)\right|^{2}
=N\sum_{f=0}^{N-1}\left|C_{j}(f)\right|^{2}\left|C_{k}(f)\right|^{2}=\frac{N^{3}}{L},
\end{split}
\end{equation}
which is independent of $j,k$. Then we have
\begin{equation}
\begin{split}
  N(MN^{2})^{2}\leq &MN^{2}\left (\frac{N^{3}}{L}+N(N-1)\theta_{A}^{2}\right )\\&+M(M-1)N^{2}\left (\frac{N^{3}}{L}+N(N-1)\theta_{C}^{2}\right ).
\end{split}
\end{equation}
Hence,
\begin{equation}
\theta_{\max}^{2}\geq \frac{N^{2}(L-1)}{L(N-1)}.
\end{equation}
Associative with the lower bound of correlation for SCS family again, we obtain the global ambiguity lower bound of SCSs as follows.
\begin{theorem}
For any $(N,M,\theta_{\max})$ SCS set $\mathcal{S}$, we have
\begin{small}
\begin{equation}
\begin{split}
\theta_{A}\geq \max\left\{N\sqrt{\frac{N-L}{L(N-1)}},N\sqrt{\frac{L-1}{L(N-1)}}\right\} \\
\end{split}
\end{equation}
\end{small}
and
\begin{small}
\begin{equation}\label{SC}
\begin{split}
\theta_{C}\geq \max\left\{\frac{N}{\sqrt{L}},N\sqrt{\frac{L-1}{L(N-1)}}\right\}.
\end{split}
\end{equation}
\end{small}
\end{theorem}

\subsection{Lower bound of DRSs with aperiodic LAZ}

Similar to the proof of  Theorem \ref{th-local} for periodic LAZ, we
can obtain the following lower bound of aperiodic ambiguity function for LAZ based on  (\ref{ZCZ})-(\ref{lasteq}). The key point is to
replace $[u_{j}(0),u_{j}(1)e^{\frac{j2\pi \nu_{i}}{N}},\cdots ,u_{j}(N-1)e^{\frac{j2\pi \nu_{i}(N-1)}{N}}]^{T}$ with $[u_{j}(0),u_{j}(1)e^{\frac{j2\pi \nu_{i}}{N}},\dots ,u_{j}(N-1)e^{\frac{j2\pi \nu_{i}(N-1)}{N}},\mathbf{0}_{1\times\left(Z_{x}-1\right)}]^{T}$ in the derivation.

\begin{theorem}\label{LAZ-bound}
For any $(N,M,\widetilde{\theta}_{\max},|\Pi|)$ unimodular DRS set $\mathcal{S}$, where $\Pi=(-Z_{x},Z_{x})\times(-Z_{y},Z_{y})$, we have
\begin{equation}\label{ALAZ-bound}
\widetilde{\theta}^{2}_{\max}\geq N^{2}\frac{MZ_{x}Z_{y}-N-Z_{x}+1}{\left(N+Z_{x}-1\right)\left(MZ_{x}-1\right)Z_{y}}.
\end{equation}
\end{theorem}

\begin{remark}
When $Z_{y}=1$, the bound of (\ref{ALAZ-bound}) is exactly  the Tang-Fan-Matsufuji bound for aperiodic correlation function derived in \cite{APbound}.
\end{remark}

\section{Two Classes of Optimal Doppler resilient Sequences}\label{sec-con}
\subsection{Characters and exponential sums over finite fields}

In this subsection, we present a brief introduction to the characters
sum and Weil bound which are important tools for the constructions of our proposed DRSs. Throughout this work, we assume that $p$ is a prime and $n$ is a positive integer. Let $q = p^n$ and $\mathbb{F}_{q}$ denote the finite field with $q$ elements. Let $Tr(\cdot)$ be the absolute trace function from $\mathbb{F}_{q}$ to $\mathbb{F}_{p}$ which is defined by
\begin{equation}
Tr(x) = x + x^{p} + \cdots + x^{p^{n-1}} , x \in \mathbb{F}_{q} .
\end{equation}

An additive character of $\mathbb{F}_{q}$ is a nonzero function $\chi$ from $\mathbb{F}_{q}$ to the set of complex numbers with absolute value of 1
such that $\chi(x + y) = \chi(x)\chi(y)$ for any pair $(x, y) \in \mathbb{F}^{2}_{q}$. For each $a\in \mathbb{F}_{q}$, the function
\begin{equation}
\chi_{a}(x)=\omega_{p}^{Tr(ax)},~x\in \mathbb{F}_{q},
\end{equation}
defines an additive character of $\mathbb{F}_{q}$, where $\omega_{p}$ is a primitive $p$-th complex root of unity. When $a = 0$, $\chi_{0}(x)=1$ for all $x \in \mathbb{F}_{q}$, and is called the trivial additive character of $\mathbb{F}_{q}$.
Let us recall the following:
\begin{lemma}\cite{Lidl}
Let $\chi$ be a nontrivial additive character of $\mathbb{F}_{q}$ and $f(x) \in \mathbb{F}_{q}[x]$ with $\operatorname{deg}(f)=d \geqslant 1$ and $\operatorname{gcd}(d, q)=1$. Then
\begin{equation}
\left|\sum_{x \in F_{q}} \chi(f(x))\right| \leqslant(d-1) \sqrt{q}.
\end{equation}
\end{lemma}

\subsection{Optimal unimodular Doppler-resilient sequences}

The concept of cubic sequence was first proposed in 1980 by Alltop to construct sequence family with good correlation properties. Although each cubic sequence does not have zero periodic autocorrelation sidelobes, the cubic sequence family meets the Welch bound in terms of its maximum periodic cross-correlation function. Formally, for a positive integer $N$, a cubic sequence family $\mathcal{U}$ is defined as follows:
\begin{Construction}\label{Con-1}
Define $\mathcal{U}=\{\mathbf{u}_{j}:0\leq j \leq N-1 \},$
where
$\mathbf{u}_{j}=\{u_{j}(0),u_{j}(1),\ldots,u_{j}(N-1)\},$
and
\begin{equation}
u_{j}(t)=\omega_{N}^{t^{3}+jt}.
\end{equation}
\end{Construction}

It is proved that the maximum periodic cross correlation magnitude of a cubic sequence family nearly achieves the Welch bound for any prime $N\geq 5$ in \cite{Alltop}. Next, we consider a generic cubic sequence, which is defined as
\begin{equation}\label{Gcubic}
u_{a,b,c}(t)=\omega_{N}^{at^{3}+bt^{2}+ct},
\end{equation}
where $a,b,c\in \mathbb{Z}_{N}$ with $a\neq 0$. 
We will show that each generic cubic sequence also possesses low cross ambiguity function, if $N$ is an odd prime number greater than 4. To proceed, let us present the following Lemma which is useful for our subsequent proof.

\begin{lemma}
Let $V_{N}(x,y)=\sum\limits_{t=0}^{N-1} \omega_{N}^{x t^{2}+yt}$, where $x,y$ are integers, $N$ is an odd integer and $g=\gcd(x,N)$, then
\begin{equation}
|V_{N}(x,y)|^{2}
=\left\{\begin{array}{ll}
Ng ;~~ & g\mid y ,\\
0 ;~~ & \rm{otherwise} .
\end{array}\right.
\end{equation}
\end{lemma}

\begin{theorem}\label{th4}
Let $N=p$ be an odd prime. The distribution of the cross-ambiguity function between $\mathbf{u}_{a_{1},b_{1},c_{1}}$ and $\mathbf{u}_{a_{2},b_{2},c_{2}}$ on family $\mathcal{U}$ is given as follows,
\begin{equation}
|AF(\tau, \nu)|=\left\{\begin{array}{ll}
p ;~~ & {\rm if}~ a_{1}=a_{2},p\mid x,p \mid y ,\\
\sqrt{p} ;~~ &{\rm if}~  a_{1}=a_{2},p\nmid x,\\
0 ;~~ &{\rm if}~  a_{1}=a_{2},p\mid x,p \nmid y,\\
\leq2\sqrt{p} ;~~ &{\rm if}~  a_{1}\neq a_{2} .
\end{array}\right.
\end{equation}
where
\begin{equation}\label{xy}
\left\{\begin{array}{ll}
x=b_{1}-3a_{2}\tau-b_{2},\\
y=c_{1}-3a_{2}\tau^{2}-2b_{2}\tau-c_{2}+\nu .
\end{array}\right.
\end{equation}
\end{theorem}
\begin{IEEEproof} Based on the definition of ambiguity function, we have
\begin{equation}
\left|AF(\tau, \nu)\right|=\left|\sum_{t=0}^{p-1}\omega_{p}^{(a_{1}-a_{2})t^{3}+xt^{2}+yt}\right|,
\end{equation}
where $x,y$ are defined in $(\ref{xy})$.

If $a_{1}=a_{2}$, this theorem holds from Lemma 3.
Otherwise, let $f(t)=(a_{1}-a_{2})t^{3}+xt^{2}+yt$ and $\chi=\chi_{1}$, based on Lemma 2, we have
\begin{equation}
\left|\sum_{t=0}^{p-1}\omega_{p}^{(a_{1}-a_{2})t^{3}+xt^{2}+yt}\right|\leq2\sqrt{p},~\text{where}~a_{1}-a_{2}\neq0.
\end{equation}
The result then follows.
\end{IEEEproof}

\begin{theorem}
The generic cubic sequence is optimal with respect to the ambiguity lower bound in \textbf{Corollary 1}.
\end{theorem}

\begin{IEEEproof} Let $N=p$, by Theorem \ref{th4} in this paper, we know
\begin{equation}
|AF_{\mathbf{u}_{a,b,c}}(\tau, \nu)|=\left\{\begin{array}{ll}
p ;~~ &  \tau=\nu=0,\\
0 ;~~ & \tau=0,\nu\neq0,\\
\sqrt{p} ;~~ & \rm{otherwise}.
\end{array}\right.
\end{equation}
Hence, the maximum ambiguity magnitude $\theta_{\max}$ of the cubic sequence set satisfies
$\theta_{\max}=\sqrt{N}.$
Then the result follows.
\end{IEEEproof}

\begin{remark}
To the best of our knowledge, the generic cubic sequence is the first known optimal unimodular sequence meeting the ambiguity lower bound in equation (\ref{eqn-bound1}). However, the lowest cross ambiguity between distinct cubic sequences is close to $2\sqrt{p}$. Whether there exists a DRS set, with more than one sequence, reaching the novel bound is still unknown.
\end{remark}

\subsection{Optimal unimodular sequences with zero ambiguity zone}

We consider certain quadratic phase CAZAC sequences which have found many applications in radar sensing, communications, coding theory, and signal processing. Specifically, for any integer $N$, a quadratic phase sequence $\mathbf{u}:\mathbb{Z}_{N}\rightarrow \mathbb{C}$ is defined by
$
u(t)= \omega_{N}^{ at^{2}+bt}, ~0\leq t\leq N-1.
$
When $N$ is odd, $\gcd(a,N)=1$ leads to the general Wiener waveform, which is a famous CAZAC sequence.

Since the quadratic phase sequence can be seen as a special case of the general cubic sequence if the coefficient of cubic term is zero. The distribution of $p$-periodic quadratic phase sequence is at most three-valued based on Theorem \ref{th4}. For any positive integer $N$, the auto-ambiguity and ZAZ of $N$-periodic quadratic phase sequence are proposed subsequently.

\begin{theorem}\label{th6}
The auto-ambiguity function of sequence $\mathbf{u}$ with $u(t)=\omega_{N}^{at^{2}+bt} $ is given as follows,
\begin{equation}
|AF_{\mathbf{u}}(\tau, \nu)|=\left\{\begin{array}{ll}
N ;~~ &  \nu\equiv2a\tau\bmod N ,\\
0 ;~~ & \rm otherwise.
\end{array}\right.
\end{equation}
\end{theorem}
\begin{IEEEproof} The above identity follows by
\begin{equation}
|AF_{\mathbf{u}}(\tau, \nu)|=\sum_{t=0}^{N-1}\omega_{N}^{(\nu-2a\tau)t}.
\end{equation}
\end{IEEEproof}

\begin{theorem}\label{th7}
Let $r=\gcd{(2a,N)}$, if $r>1$, the maximum zero auto-ambiguity zone of quadratic phase sequences is given below:
\begin{equation}
\Pi=\left(-\frac{N}{r},\frac{N}{r}\right)\times \left(-r,r\right),
\end{equation}
which is optimal with respect to the ambiguity lower bound in equation $(5)$ .
\end{theorem}
\begin{IEEEproof} By Theorem \ref{th6}, we observe the peak points of auto-ambiguity function near the origin are $(\frac{N}{r},0)$ and $\left(h,r\right)$, where $\frac{2a}{r}\cdot h\equiv1\bmod \frac{N}{r}$. Since $r\mid y$ for any peak points $(x,y)$, there does not exist non-zero ambiguity function value in the region $\Pi$. Note that the area of $\Pi$ is $N$ and this completes the proof.
\end{IEEEproof}

Theorem \ref{th6} and Theorem \ref{th7} indicate that each quadratic phase sequence has a zero auto-ambiguity zone. Next, we propose a construction of quadratic phase sequence family with optimal zero ambiguity zone.

\begin{Construction}
With the same notations as above, let $b_{i}=i\lfloor \frac{N}{M}\rfloor$,  define a sequence family $\mathcal{S}$ with integer $M$ as $
\mathcal{S}=\{\mathbf{s}_{i}:0\leq i\leq M-1\},
$
where
${\mathbf{s}_{i}}=[s_i(0),s_i(1),\cdots,s_i(N-1)],$
is given by
\begin{equation}
s_{i}(t)=\omega_{N}^{at^{2}+b_{i}t},0\leq t\leq N-1.
\end{equation}
\end{Construction}
Then we obtain an $(N,M,0,4\lfloor\frac{N}{M}\rfloor)$ unimodular DRS set $\mathcal{S}$, which is optimal with respect to the ambiguity lower bound in equation $(\ref{LAZ-bound})$ if $M\mid N$.
\begin{IEEEproof}
By Theorem 7, we observe that the zero ambiguity zone of $\mathcal{S}$ is in the area of
\begin{equation}
\Pi=\left(-\frac{N}{r},\frac{N}{r}\right)\times \left(-r,r\right).
\end{equation}
Similar to Theorem \ref{th6}, the peak points $(\tau,\nu)$ of cross-ambiguity between ${\mathbf{s}_{i}}$ and ${\mathbf{s}_{j}}$ satisfy $v\equiv2a\tau-b_{i}+b_{j}\mod N$. Hence, no peak points exist in the area of $\left(-\frac{\min_{i\neq j}\{b_{i}-b_{j}\}}{r},\frac{\min_{i\neq j}\{b_{i}-b_{j}\}}{r}\right)\times (-r,r)$. If $M\mid N$, the area of $\Pi$ is $\frac{4N}{M}$ and this completes the proof.
\end{IEEEproof}

\section{Optimal SCSs with Doppler Resilience }

In this section, we construct an optimal SCS with respect to the bound in Theorem 3 using cyclic difference sets. First, we give the definition of cyclic difference set as follows.

For any subset $\mathcal{D} = \{d_{0}, d_{1},\ldots,d_{n-1}\} \in \mathbb{Z}_{N}$, the difference function of $\mathcal{D}$ is defined as
$
d_{\mathcal{D}}(\varepsilon) = |(\varepsilon + \mathcal{D})\cap \mathcal{D}|,~~ \varepsilon \in \mathbb{Z}_{N} .
$
Then $\mathcal{D}$ is said to be an $(N, n, d)$ cyclic difference set if and only if $d_{\mathcal{D}}(\varepsilon)$ takes on the value $d$ for $N-1$ times when $\varepsilon$ ranges over the nonzero elements of $\mathbb{Z}_{N} $.

\begin{Construction}
Let $\mathcal{D}$ be an $(N,n,1)$ cyclic difference set, a frequency domain sequence $\mathbf{C}$ with $N-n$ spectrum holes is defined as
\begin{equation}
C(f)=\left\{\begin{array}{ll}
\sqrt{\frac{N}{n}}(-1)^{f};~~ & f\in \mathcal{D},\\
0 ;~~ & \rm{otherwise}.
\end{array}\right.
\end{equation}
Then the time domain dual sequence $\mathbf{c}$ is given as
\begin{equation}
c(t)=\frac{1}{\sqrt{N}}\sum_{f=0}^{N-1}C(f)e^{j2\pi ft/N}.
\end{equation}
\end{Construction}

\begin{lemma}
The distribution of auto-ambiguity function value for the time domain dual sequence $\mathbf{c}$ is given as follows,
\begin{equation}
|AF_{\mathbf{c}}(\tau, \nu)|=\left\{\begin{array}{ll}
N ;~~ & \tau=0,\nu=0 ,\\
\frac{N\sqrt{n-1}}{n} ;~~ & \tau\neq0,\nu=0 ,\\
\frac{N}{n} ;~~ & \rm{otherwise} .
\end{array}\right.
\end{equation}
\end{lemma}

\begin{theorem}
The normalized time domain dual sequence $\mathbf{c}$ in Construction 3 is an optimal SCS with respect to the bound in \textbf{Theorem 3}.
\end{theorem}

\begin{IEEEproof} Based on the property of cyclic difference set, we have $n(n-1)=d(N-1)=N-1$. Therefore, the lower bound of Theorem 3 reduces to $\max\{\frac{N\sqrt{n-1}}{n},\frac{N}{n}\}$, and this completes the proof.
\end{IEEEproof}

Finally, we present a simple construction of SCS family based on orthogonal sequence family whose maximum ambiguity value achieves the lower bound in Theorem 2.

\begin{Construction}
Let $\mathcal{O}=\{\mathbf{O}_{i}:i=0,1,\ldots,N-1\}$ is an orthogonal unimodular sequence family with period $N$ in the frequency domain.
Let $k$ be an integer, define a sequence family with period $kN$ as
$
\{\mathbf{T}_{i}:i=0,1,\ldots,N-1\},
$
where
${\mathbf{T}_{i}}=[T_i(0),T_i(1),\cdots,T_i(kN-1)]$
is given by
\begin{equation}
T_{i}(f)=\left\{
             \begin{array}{ll}
               \sqrt{k}O_{i}(\frac{f}{k}); & \rm{if}~f\equiv 0 \bmod k ,\\
               0; & \rm{otherwise}.
             \end{array}
           \right.
\end{equation}
\end{Construction}
Let $\mathbf{t}_{i}$ be the time domain dual sequence of $\mathbf{T}_{i}$, and $\mathcal{T}=\{\mathbf{t}_{i}:i=0,1,\ldots,N-1\}$.
Then we obtain a $(kN,N,0,4k)$ Doppler-resilient SCS set $\mathcal{T}$ with zero ambiguity zone, which is optimal with respect to the bound in Theorem 2.
\begin{IEEEproof}
Based on the definition of ambiguity function, we have
\begin{equation}
\begin{split}
|AF_{\mathbf{t}_{i},\mathbf{t}_{j}}(&\tau, \nu)|=\left\{\begin{array}{ll}
k\sum\limits_{f=0}^{N-1}  O_{i}(f-\frac{\nu}{k})O_{j}^{*}(f)\omega_{N}^{ f\tau} ;~~ &  k\mid \nu ,\\
0 ;~~ & \rm{otherwise}.
\end{array}\right.
\end{split}
\end{equation}
Hence, $Z_{y}\geq k$.
Since $\mathcal{O}$ is an orthogonal sequence family, we have
\begin{equation}
|AF_{\mathbf{t}_{i},\mathbf{t}_{j}}(\tau, 0)|=0,~\text{for}~i\neq j,
\end{equation}
so $Z_{x}\geq 1$.
Therefore, $NZ_{x}Z_{y}\geq kN$. By recalling the Theorem 2 completes the proof.
\end{IEEEproof}

\section{Examples and potential applications}

\subsection{Examples of optimal DRSs with periodic ZAZ and LAZ}
In this section, we present four examples to illustrate our proposed constructions.

\begin{example}\label{ex1}
Let $N=32$, $M=2$ and $0\leq t\leq31$, define
\begin{equation}
u(t)=e^{\frac{j2\pi (2t^{2})}{32}}~~\rm{and}~~v(t)=e^{\frac{j2\pi (2t^{2}+16t)}{32}}.
\end{equation}

The ambiguity function value of  $\mathcal{S}=\{\mathbf{u},\mathbf{v}\}$ is given in Fig. \ref{f4}, where each point refers to an ambiguity peak point and zero otherwise.
The ZAZ of the sequence family is $(-4,4)\times(-4,4)$, which is optimal with respect to the bound in Theorem 1.
\begin{figure}[!htbp]
  \centering
  \includegraphics[width=9.5cm]{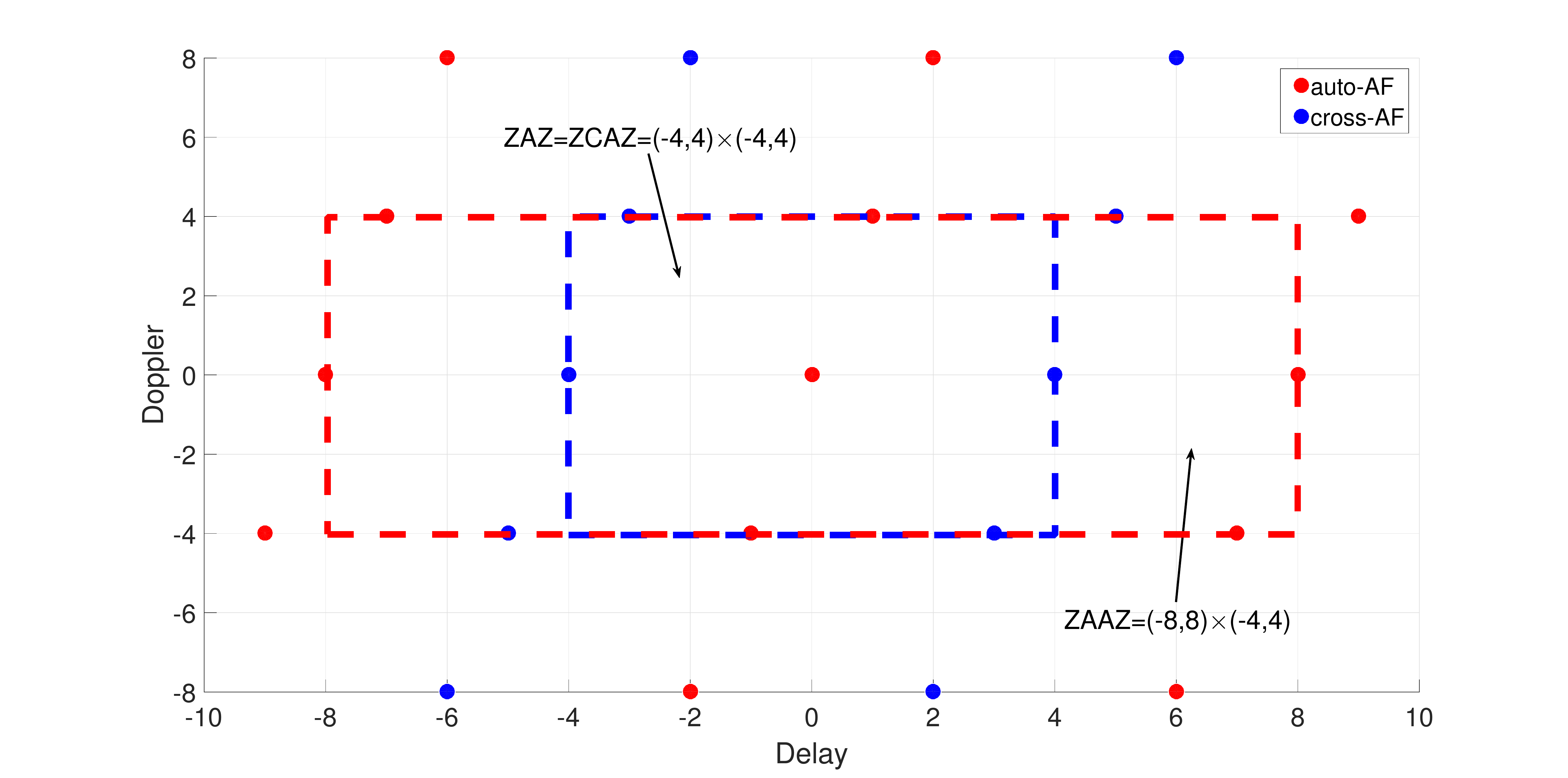}
  \caption{ ZAZ of the sequence family $\mathcal{S}$ in Example \ref{ex1}.}
  \label{f4}
\end{figure}
\end{example}

\begin{example}\label{ex2}
Let $p=31$ and $0\leq t\leq30$, define
\begin{equation}
u(t)=e^{\frac{j2\pi t^{3}}{31}}~~\rm{and}~~v(t)=e^{\frac{j2\pi (t^{3}+15t)}{31}},
\end{equation}
then the magnitude of auto-ambiguity function value of sequence $\mathbf{u}$ is the same as $\mathbf{v}$, which is presented in Fig. \ref{f5}.

\begin{figure}[!htbp]
  \centering
  \includegraphics[width=9.5cm]{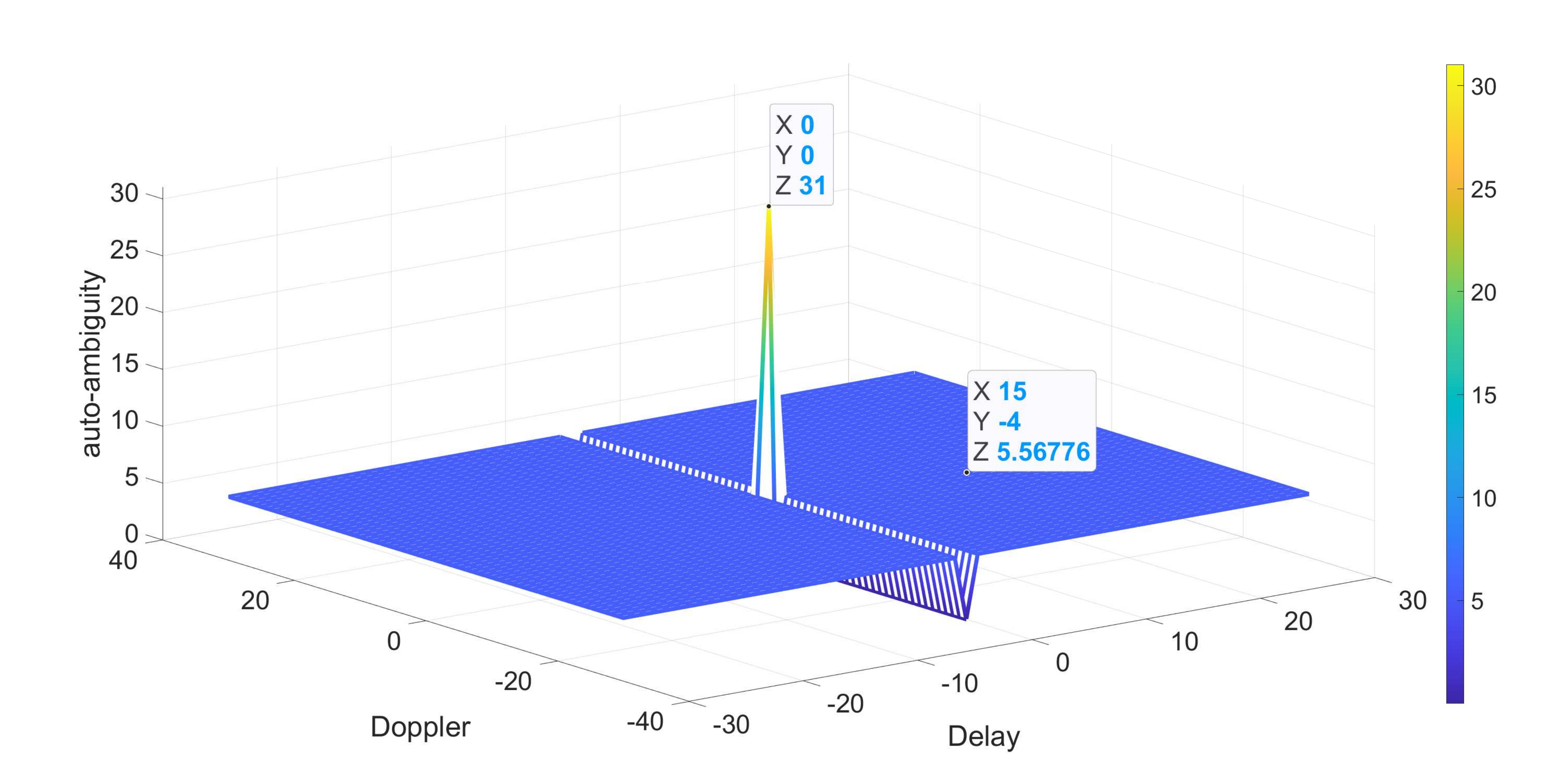}
  \caption{ Auto-ambiguity function  of sequence $\mathbf{u}$ in Example \ref{ex2}.}
  \label{f5}
\end{figure}
\begin{figure}[!htbp]
  \centering
  \includegraphics[width=9.5cm]{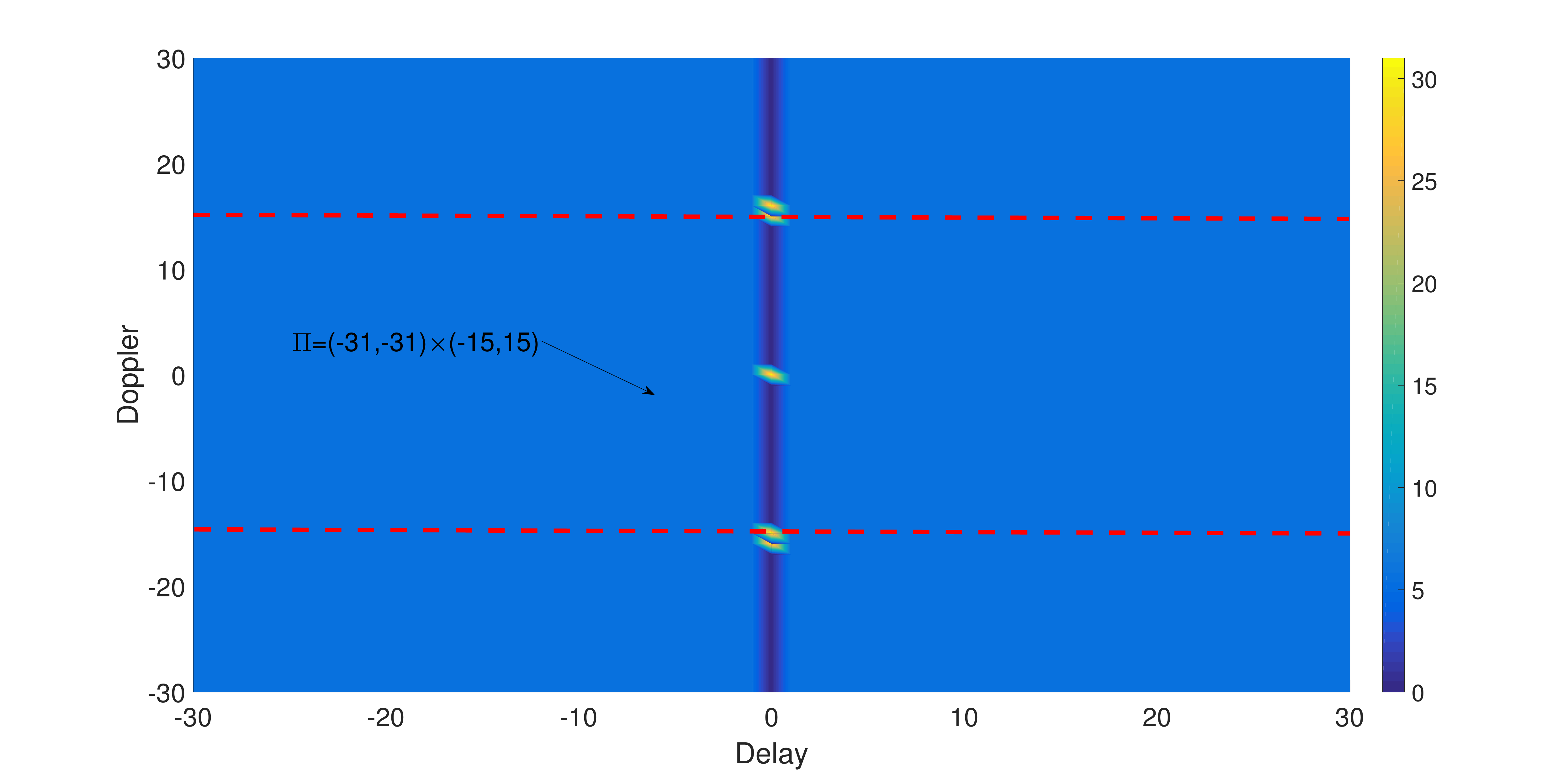}
  \caption{LAZ of sequence set $\mathcal{S}$ in Example \ref{ex2}.}
  \label{f6}
\end{figure}
\end{example}

Specifically,
\begin{equation}
\begin{split}
|AF_{\mathbf{u}}(\tau, \nu)|=\left\{\begin{array}{ll}
31 ;~~ &  \tau=\nu=0 ,\\
0 ;~~ & \tau=0,\nu\neq0,\\
\sqrt{31} ;~~ & \rm{otherwise},
\end{array}\right.
\end{split}
\end{equation}
which is optimal with respect to (\ref{eqn-bound1}) in Corollary 1.
In addition, sequences $\mathbf{u}$ and $\mathbf{v}$ form a $(31,2,\sqrt{31},(-31,31)\times(-15,15))$ DRS set $\mathcal{S}$, the low ambiguity zone is presented in Fig. \ref{f6}.

\begin{example}\label{ex3}
Let $D=\{4,5,8,10\}$ be a $(13,4,1)$ cyclic difference set with $N=13$, then
\begin{equation}
C(f)=\left[ 0,0,0,0,\sqrt{\frac{13}{4}},-\sqrt{\frac{13}{4}},0,0,\sqrt{\frac{13}{4}},0,\sqrt{\frac{13}{4}},0,0\right].
\end{equation}
The magnitude of auto-ambiguity function value of corresponding time domain sequence $\mathbf{c}$ is shown in Fig. \ref{f7}. Specially,
\begin{equation}
|AF_{\mathbf{c}}(\tau, \nu)|=\left\{\begin{array}{ll}
13 ;~~ & \tau=0,\nu=0 ,\\
13\sqrt{3}/4 ;~~ & \tau\neq0,\nu=0 ,\\
13/4 ;~~ & \rm{otherwise} ,
\end{array}\right.
\end{equation}
which is optimal with respect to the bound in Theorem 3.
\begin{figure}[h]
\vspace{-0.35cm}  
\setlength{\belowcaptionskip}{-0.4cm}   
  \centering
  \includegraphics[width=9cm]{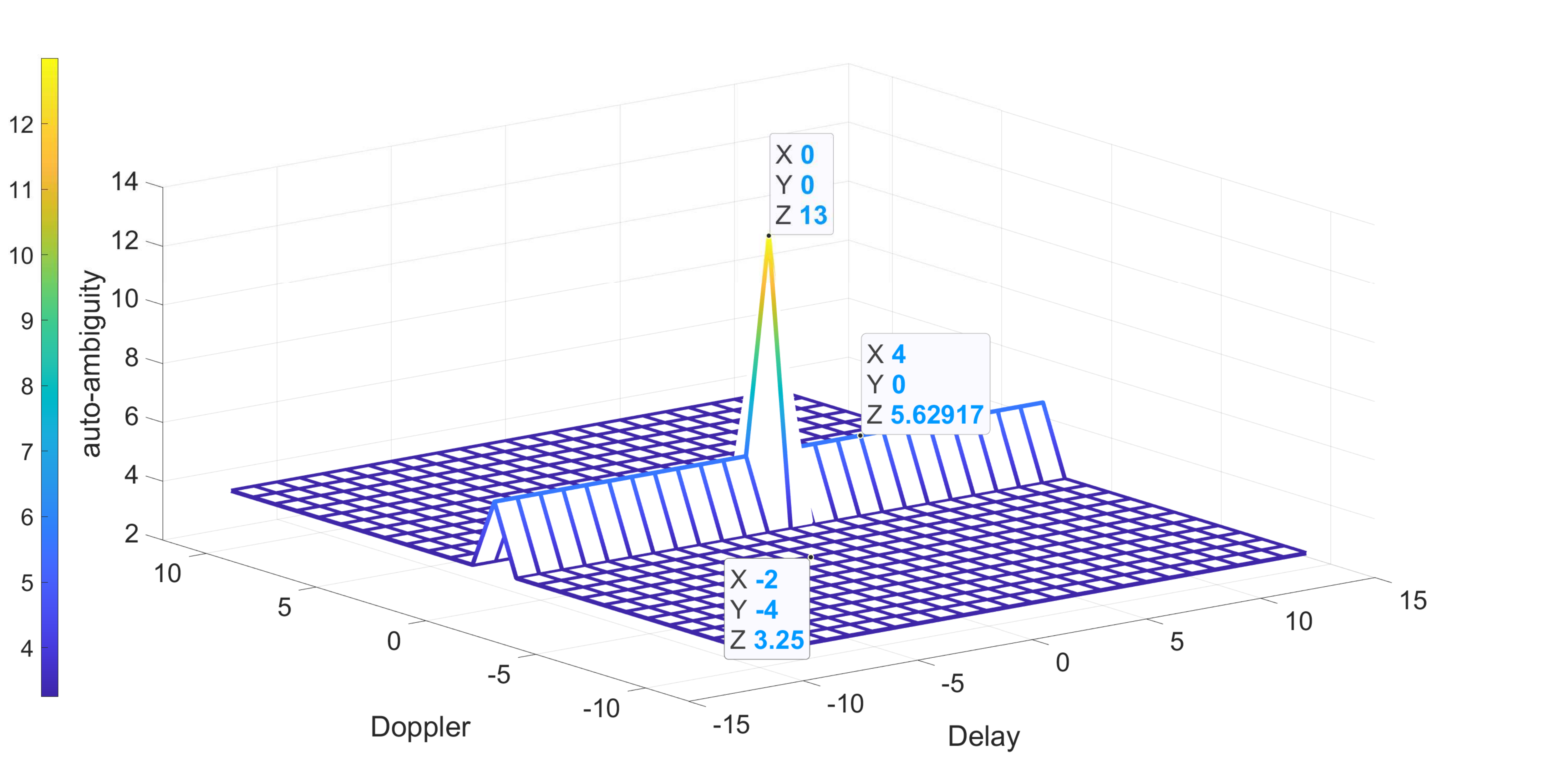}
  \caption{ Auto-ambiguity function  of sequence $\mathbf{u}$ in Example \ref{ex3}.}
\label{f7}
\end{figure}
\end{example}

\begin{example}\label{ex4}
Let $N=8,M=2$, define
\begin{equation}
\mathbf{T_{1}}=\sqrt{2}\cdot[ 1,0,1,0,1,0,1,0]
\end{equation}
and
\begin{equation}
\mathbf{T_{2}}=\sqrt{2}\cdot[ 1,0,-1,0,1,0,-1,0].
\end{equation}
The magnitude of auto-ambiguity function value of corresponding time domain sequence family $\mathcal{T}$ is shown in Fig. \ref{f8}. The ZAZ is $(-2,2)\times(-2,2)$ which is optimal with respect to the bound of Theorem 2.
\begin{figure}[h]
\vspace{-0.35cm}  
  \centering
  \includegraphics[width=9cm]{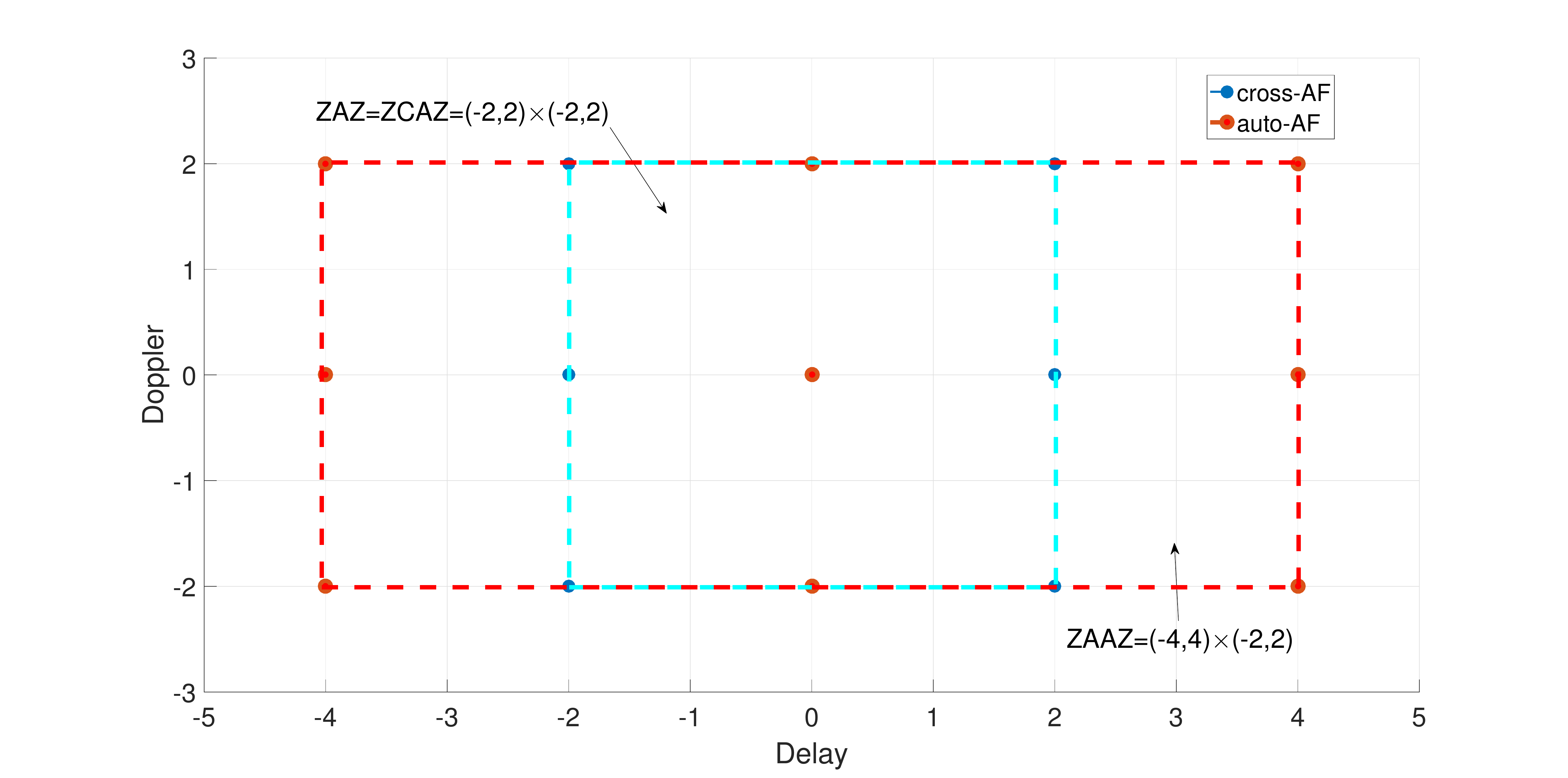}
  \caption{ ZAZ of the sequence family $\mathcal{T}$ in Example \ref{ex4}.}
\label{f8}
\end{figure}
\end{example}

\subsection{Examples of DRSs with aperiodic LAZ}
In general, it is more difficult to study the aperiodic AF than periodic one, and is very challenging to obtain DRSs with aperiodic AF via algebraic and combinatorial method.
However, one can obtain DRSs with aperiodic LAZ based on some optimization algorithms as reported in \cite{Cui,Jing,Wang2021}. In the following, to illustrate the existence of DRSs with aperiodic LAZ, we give an example based on the heuristic algorithm in \cite{Zhang21}.

\begin{example}\label{ex5}
Let $\mathbf{s}$ be a binary sequence of length 128 given by
\begin{equation}
\begin{split}
\mathbf{s}=&[-+--++-+-++++++---+--++\\
           &-+---++-+-++++++--++-+-\\
           &---+-+++--+-----++-+-++\\
           &++--+---+-+++--+-+++++-\\
           &+--+++-+++--+-+++++--+-\\
           &+++-+++++---+],
\end{split}
\end{equation}
where $+$ and $-$ denote $1$ and $-1$ respectively.  The magnitude of auto-ambiguity function value of $\mathbf{s}$ is shown in Fig. \ref{f9} and the aperiodic LAZ is $(-4,4)\times(-4,4)$.

\begin{figure}
\vspace{-0.35cm}  
\setlength{\belowcaptionskip}{-0.6cm}   
  \centering
  \includegraphics[width=9cm]{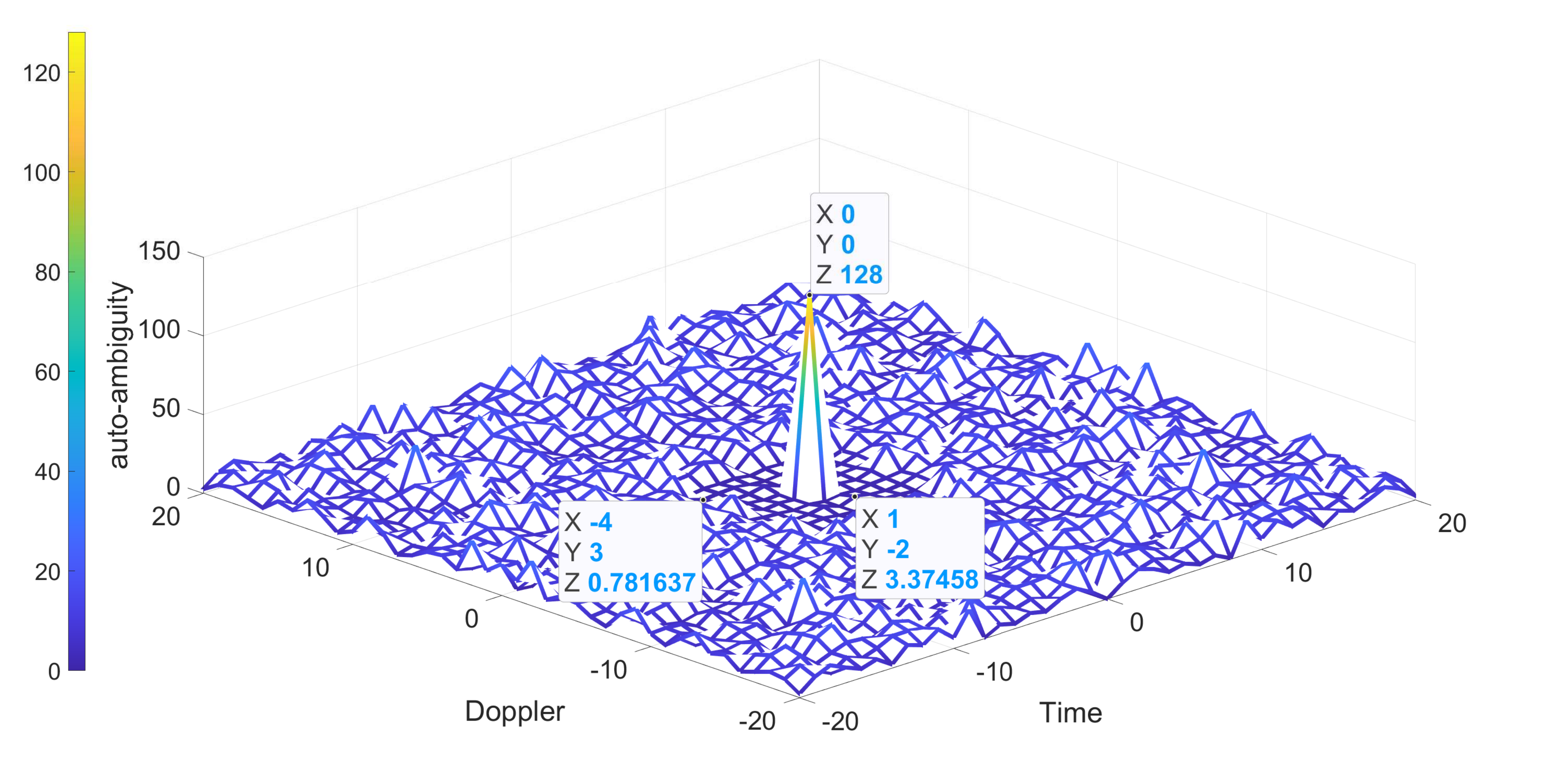}
  \caption{Aperiod LAZ of the sequence $\mathbf{s}$ in Example \ref{ex5}.}
\label{f9}
\end{figure}
\end{example}

\subsection{Potential applications}
There are three types of sensing code signal design, corresponding to three different ISAC systems \cite{Ma2020,Cui2021}, i.e.
\begin{enumerate}
  \item Communication-centric design (communications signal-based ), such as OFDM/OTFS signals, spread-spectrum signals, Protocol-oriented design. The performance metric for such signal design include correlation function, PAPR, SINR, etc.
  \item Radar-centric design (radar signal-based), such as CW signals, Pulsed signals, FMCW, SFCW, Phase-coded radar signals, PTM-GCP, NS-GCP, etc. The performance metric for radar-centric signal design is mainly AF, other metric include, mutual information (MI), CRLB, etc.
  \item Joint design normally involves solving a joint optimization problem to balance radar and communication performance, relying on prior channel state knowledge. In joint design, the objective function and the constraints can use various metric and their combinations, for communications, and for radar sensing.
\end{enumerate}

    The DRSs with periodic/aperiodic LAZ/ZAZ characteristics discussed in this paper can be applied to radar-centric ISAC systems, which can embed data bits into radar signals by modifying the radar signal to incorporate digital modulations, or by index modulation (IM) to send data bits via the indices of certain radar parameters (MIMO, FAR), although the data rate in this ISAC system is normally limited. For this radar-centric design, it is very important for the radar sensing signal to have superior ambiguity function. In many cases, the maximum tolerable Doppler shift is much smaller than the sensing signal bandwidth; further, it is not necessary to consider the maximum unambiguous distance range. Thus one may only need to consider a small area of interests near-origin zone, i.e. the low or zero ambiguity zone (LAZ/ZAZ). In fact, the proposed new LAZ/ZAZ concepts, the derived theoretical bounds and the presented DRSs with periodic LAZ/ZAZ are the main contributions of this paper. The discussed DRSs with periodic/aperiodic LAZ/ZAZ characteristics can find potential applications in ISAC systems under the frameworks developed in \cite{Hassanien2019,Ted2018,Kum2018,Duggal2021}.

The first potential application is to implement the dual functions of radar and communication by embedding information in our obtained sequences. In \cite{Hassanien2019} and \cite{Ted2018}, for example, a distinctive Gold/Kasami sequence is sent to represent a communication symbol through code keying, whereas the resultant pseudo-random communication signals are used as radar waveform. It is pointed out in \cite{Hassanien2019} and \cite{Ted2018} that the data rate and the sensing performance are determined by the size and periodic AF of the employed sequence family, respectively. Under this framework, the sequence family constructed in Section V can also be used to implement the dual functions of radar and communication. Specially, we give a simple comparison of the sequence family $\mathcal{U}$ generated by Construction 1 with the Gold/Kasami sequence families.
For certain  sequence length $N$, the sizes of Gold sequence family, Kasami sequence family and the proposed family $\mathcal{U}$ are $N+1,\sqrt{N+1}$ and $N-1$, respectively\footnote{$N=2^{2k+1}-1$ for Gold sequence family, $N=2^{2k}-1$ for Kasami sequence family, and $N$ is a prime for $\mathcal{U}$. Hence, we choose approximate period for comparison.}.
Therefore, compared to the Gold sequence family (resp., Kasami sequence family) with almost the same length, the proposed sequence family  $\mathcal{U}$ can support almost the same (resp., larger) date rate when they are employed under the framework developed in \cite{Ted2018}. Besides, the proposed sequence family $\mathcal{U}$ enjoys excellent periodic AF, thus leading to potential improved sensing performance.

A second potential application is for enhanced preamble and sensing waveform in future
vehicular communication-radar systems, under the frameworks developed in \cite{Kum2018,Duggal2021}. For communications, such sequences can be used as preambles to achieve the tasks of synchronization and channel estimation due to their good correlation properties. On the other hand, they can also be employed to construct radar waveforms for improving sensing, thanks to their very low AF sidelobe in a local region.
Following this framework, our proposed sequences can also be used to perform the dual functions of radar and communications.

%

\section{Conclusion}
In this paper, in order to meet very low ambiguity function requirement within a small area of interest defined by the certain Doppler frequency and delay, and to design the desired optimal signals for sensing and communication systems, a new concept called low ambiguity zone, as well as zero ambiguity zone, was proposed. Based on the new concept, we derived four lower bounds on periodic LAZ/ZAZ of unimodular DRSs with and without spectral constraints. These bounds may be used as benchmarks to measure the Doppler resilience of unimodular sequences. Also, we presented four types of optimal DRSs with respect to a collection of the proposed bounds. It should be noted that our ambiguity lower bounds of SCSs are derived with the assumption that all the sequences share the identical spectral hole constraint $\Omega$. New ambiguity lower bounds of SCSs may be possible when different spectral hole constraints are to be considered. In addition, it is also interesting, although much more difficult, to investigate aperiodic ambiguity lower bounds and the associated optimal DRSs.



\section{Appendix}

\subsection{Proof of Lemma 1}
If $\mathbf{u}_{j}$ and $\mathbf{u}_{k}$ are polyphase sequences, for arbitrary time shift $\tau$, we have the ambiguity function satisfying
\begin{equation}
\begin{split}
&\sum_{\nu=0}^{N-1}\left|AF_{\mathbf{u}_{j},\mathbf{u}_{k}}(\tau, \nu)\right|^{2}\\
=&\sum_{\nu=0}^{N-1}\sum_{t=0}^{N-1}u_{j}(t) u^{*}_{k}(t+\tau) e^{\frac{j2\pi \nu t}{N}}\sum_{s=0}^{N-1}u^{*}_{j}(s) u_{k}(s+\tau) e^{-\frac{j2\pi \nu s}{N}}\\
=&\sum_{t=0}^{N-1}\sum_{s=0}^{N-1}u_{j}(t) u^{*}_{k}(t+\tau)u^{*}_{j}(s) u_{k}(s+\tau)\sum_{\nu=0}^{N-1}e^{\frac{j2\pi \nu(t-s)}{N}}\\
=&N\sum_{t=0}^{N-1}| u_{j}(t)|^{2} | u_{k}(t+\tau)|^{2}=N^{2}.
\end{split}
\end{equation}
Since
$
AF_{\mathbf{u}_{j}}(0,0)=N,
$
we have
\begin{equation}
\left|AF_{\mathbf{u}_{j}}(0, \nu)\right|=0,~\text{for}~\nu\neq 0.
\end{equation}

\subsection{Proof of Corollary 1}
If $Z_{x}=Z_{y}=N$, by denoting $\mathbf{U}_{(Z_{x},Z_{y})}$ as $\mathbf{U}$, we have
\begin{equation}\label{glob}
\parallel \mathbf{U}\mathbf{U}^{H} \parallel_{F}^{2}=N(MN^2)^{2}.
\end{equation}
Based on the definition of ambiguity magnitude of DRS family and (\ref{polyphase}), (\ref{ZCZ}), (\ref{bound1}), we have
\begin{equation}
MN^{4}+MN^{2}(N^{2}-N)\theta_{A}^{2}+MN^{2}(MN^{2}-N^{2})\theta_{C}^{2}\geq N(MN^2)^{2}.
\end{equation}
Hence,
$
\theta_{\max}\geq \sqrt{N}.
$
The rest of the proof is similar to the above, so we omit it.

\subsection{Proof of Lemma 3}
By definition
\begin{equation}
\begin{split}
|V_{N}(x,y)|^{2}&=\sum\limits\limits_{t=0}^{N-1} \omega_{N}^{x t^{2}+yt}\sum\limits_{s=0}^{N-1} \omega_{N}^{-x s^{2}-ys}\\
&=\sum\limits_{t=0}^{N-1}\sum\limits_{s=0}^{N-1} \omega_{N}^{(t-s)(x(t+s)+y)}\\
&=\sum\limits_{k=0}^{N-1}\sum\limits_{s=0}^{N-1} \omega_{N}^{k(x(2s+k)+y)}\\
&=\sum\limits_{k=0}^{N-1}\omega_{N}^{xk^{2}+yk}\sum\limits_{s=0}^{N-1} \omega_{N}^{2kxs}.
\end{split}
\end{equation}
If $N$ is odd, then
\begin{equation}
\sum\limits_{s=0}^{N-1} \omega_{N}^{2kxs}=\left\{\begin{array}{ll}
N ;~~ & N\mid kx ,\\
0 ;~~ & \text{otherwise} .
\end{array}\right.
\end{equation}
Hence, let $g=\gcd(x,N)$, we have
\begin{equation}
\begin{split}
|V_{N}(x,y)|^{2}
&=N\sum\limits_{N\mid kx}\omega_{N}^{xk^{2}+yk}
=N\sum\limits_{e=0}^{g-1}\omega_{N}^{\frac{yeN}{g}}\\
=N\sum\limits_{e=0}^{g-1}\omega_{g}^{ye}
&=\left\{\begin{array}{ll}
Ng ;~~ & g\mid y ,\\
0 ;~~ & otherwise .
\end{array}\right.
\end{split}
\end{equation}
This completes the proof.

\subsection{Proof of Lemma 4}
By the definition of the ambiguity function in equation $(\ref{bound1})$ ,
if $\nu\neq0$, we have
\begin{equation}
|AF_{\mathbf{c}}(\tau, \nu)|=\left|\sum\limits_{f\in\mathcal{D}\cap \mathcal{D}+\nu } C^{*}(f) C(f-\nu) e^{-j 2 \pi f \tau/N}\right|
\end{equation}
and
\begin{equation}
|AF_{\mathbf{c}}(\tau, 0)|=\frac{N}{n}\left|\sum\limits_{f\in\mathcal{D}}  e^{-j 2 \pi f \tau/N}\right|.
\end{equation}
Since $\mathcal{D}$ is an $(N,n,1)$ cyclic difference set, we have $\left|\mathcal{D}\cap \left(\mathcal{D}+v\right)\right|=1$, for $v\neq0$, then
$
|AF_{\mathbf{c}}(\tau, \nu)|=\frac{N}{n}.
$
If $v=0$, we have
\begin{equation}
|AF_{\mathbf{c}}(\tau, 0)|^{2}=\frac{N}{n}\left|\sum\limits_{f,f'\in\mathcal{D}}  e^{-j 2 \pi (f-f') \tau/N}\right|.
\end{equation}
Thus, we have $|AF_{\mathbf{c}}(\tau, 0)|=\frac{N\sqrt{n-1}}{n}$, for $\tau\neq0$, based on the properties of the cyclic difference set.
Then the result follows.

\bibliographystyle{latex8}
\bibliography{latex8}

\begin{IEEEbiography}[{\includegraphics[width=1in,height=1.25in,clip,keepaspectratio]{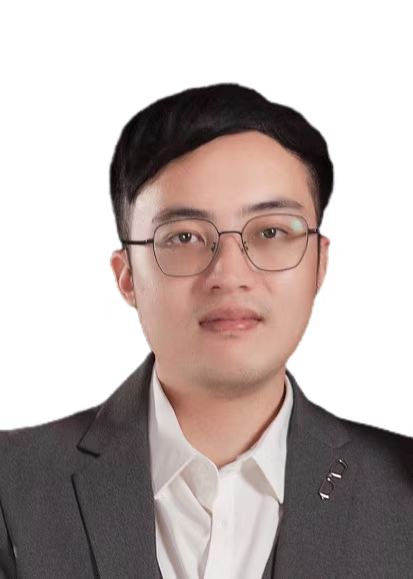}}]
{Zhifan Ye}
received the B.S. and M.S. degrees in mathematics
from Fujian Normal University, Fuzhou, China, in 2013 and 2016, respectively.
He is currently pursuing the Ph.D. degree with the School of Mathematics from Southwest Jiaotong University, Chengdu, China. His research interests include sequence design, sensing and communication.

\end{IEEEbiography}

\begin{IEEEbiography}[{\includegraphics[width=1in,height=1.25in,clip,keepaspectratio]{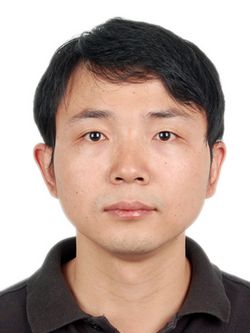}}]
{Zhengchun Zhou}
received the B.S. and M.S. degrees in mathematics
and the Ph.D. degree in information security from Southwest Jiaotong
University, Chengdu, China, in 2001, 2004, and 2010, respectively.
From 2012 to 2013, he was a postdoctoral member in the Department
of Computer Science and Engineering, the Hong Kong University
of Science and Technology. From 2013 to 2014, he was a research
associate in the Department of Computer Science and Engineering,
the Hong Kong University of Science and Technology. Since
2001, he has been in the Department of Mathematics, Southwest
Jiaotong University, where he is currently a professor. His research
interests include sequence design, Boolean function, coding theory,
and compressed sensing. He is an associated editor of Advances in
Mathematics of Communications and IEICE Transactions on Fundamentals,
and was a Guest Editor for special issues of Cryptography
and Communications.

Dr. Zhou was the recipient of the National excellent Doctoral
Dissertation award in 2013 (China).
\end{IEEEbiography}

\begin{IEEEbiography}[{\includegraphics[width=1in,height=1.25in,clip,keepaspectratio]{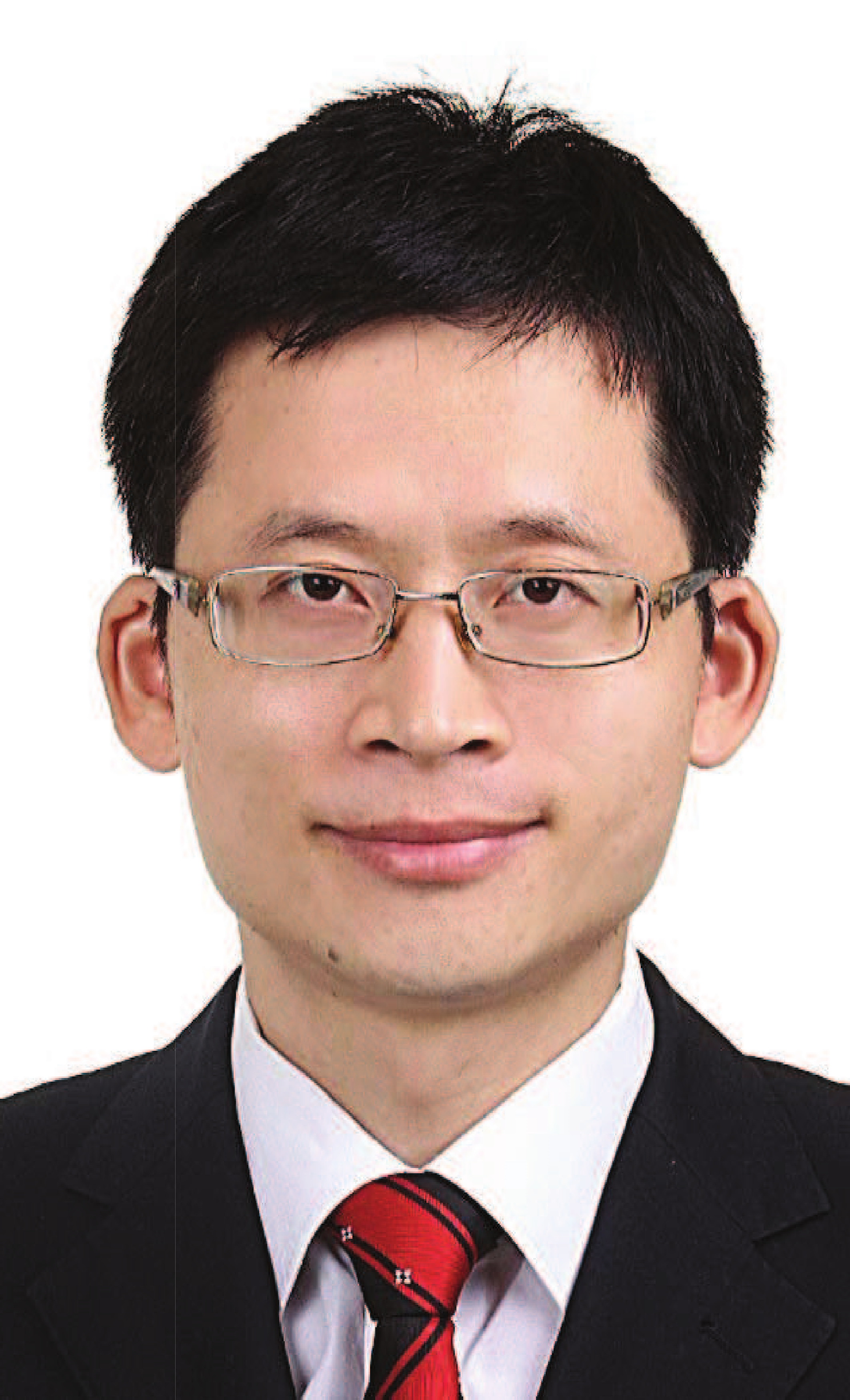}}]
{Zilong Liu}
received his PhD (2014) from School of Electrical and Electronic Engineering, Nanyang Technological University (NTU), Master Degree (2007) in the Department of Electronic Engineering from Tsinghua University, and Bachelor Degree (2004) in the School of Electronics and Information Engineering from Huazhong University of Science and Technology (HUST). He is a Lecturer at the School of Computer Science and Electronic Engineering, University of Essex. From January 2018 to November 2019, he was a Senior Research Fellow at the Institute for Communication Systems (ICS), Home of the 5G Innovation Centre (5GIC), University of Surrey. Prior to his career in UK, he spent 9.5 years in the School of Electrical and Electronic Engineering, Nanyang Technological University (NTU), Singapore, first as a Research Associate (since July 2008) and then a Research Fellow (since November 2014). He is generally interested in coding and signal processing for various communication systems. Details of his research can be found at: https://sites.google.com/site/zilongliu2357.
\end{IEEEbiography}

\begin{IEEEbiography}[{\includegraphics[width=1in,height=1.25in,clip,keepaspectratio]{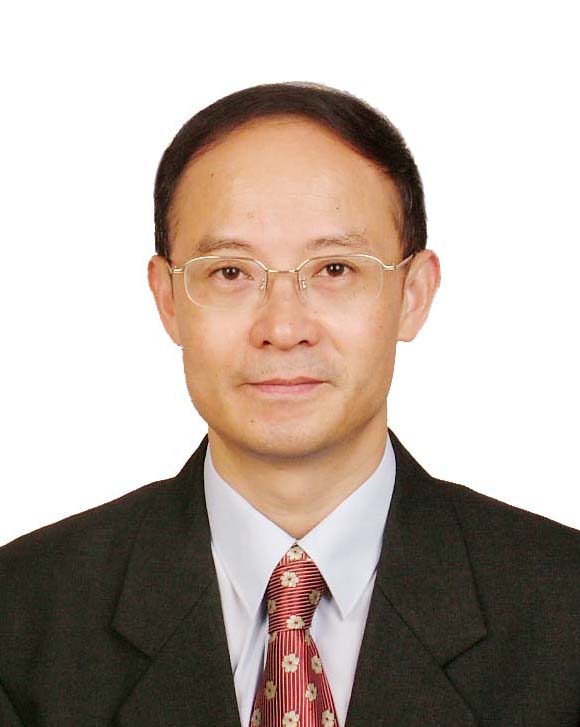}}]
{Pingzhi Fan (M'93-SM'99-F'15)}
received his MSc degree in computer science from the Southwest Jiaotong University, China, in 1987, and PhD degree in Electronic Engineering from the Hull University, UK, in 1994. He is currently a distinguished professor and director of the institute of mobile communications, Southwest Jiaotong University, China, and a visiting professor of Leeds University, UK (1997-), a guest professor of Shanghai Jiaotong University (1999-). He is a recipient of the UK ORS Award (1992), the NSFC Outstanding Young Scientist Award (1998), IEEE VTS Jack Neubauer Memorial Award (2018), and 2018 IEEE SPS Signal Processing Letters Best Paper Award. His current research interests include vehicular communications, massive multiple access and coding techniques, etc. He served as general chair or TPC chair of a number of international conferences including VTC' 2016 Spring, IWSDA'2019, ITW'2018 etc. He is the founding chair of IEEE Chengdu (CD) Section, IEEE VTS BJ Chapter and IEEE ComSoc CD Chapter. He also served as an EXCOM member of IEEE Region 10, IET(IEE) Council and IET Asia Pacific Region. He has published over 300 international journal papers and 8 books (incl. edited), and is the inventor of 25 granted patents. He is an IEEE VTS Distinguished Lecturer (2015-2019), and a fellow of IEEE, IET, CIE and CIC.
\end{IEEEbiography}

\begin{IEEEbiography}[{\includegraphics[width=1in,height=1.25in,clip,keepaspectratio]{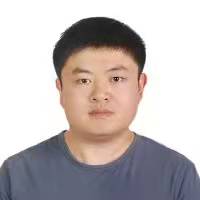}}]
{Xianfu Lei}
received his Ph.D from Southwest Jiaotong University in 2012. He has been an Associate Professor with the School of Information Science and Technology at Southwest Jiaotong University since 2015. From 2012 to 2014, he worked as a research fellow in the Department of Electrical and Computer Engineering at Utah State University. Dr Lei's research interests are 5G/6G networks, cooperative and energy harvesting networks and physical-layer security. He has been serving as an Area Editor for IEEE Communications Letters and an Associate Editor for IEEE Wireless Communications Letters and IEEE Transactions on Communications. He served as Senior/Associate Editor for IEEE Communications Letters from 2014-2019. He received the best paper award in IEEE/CIC ICCC2020, the best paper award in WCSP2018, the WCSP 10-Year Anniversary Excellent Paper Award, IEEE Communications Letters Exemplary Editor 2019, and Natural Science Award of China Institute of Communications (2019).
\end{IEEEbiography}

\begin{IEEEbiography}[{\includegraphics[width=1in,height=1.25in,clip,keepaspectratio]{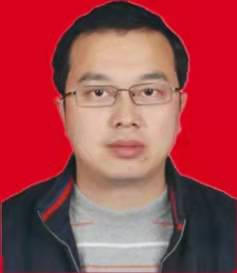}}]
{Xiaohu Tang}(Senior Member, IEEE)
 received the B.S. degree in applied
mathematics from Northwestern Polytechnical University, Xi'an, China,
in 1992, the M.S. degree in applied mathematics from Sichuan University,
Chengdu, China, in 1995, and the Ph.D. degree in electronic engineering
from Southwest Jiaotong University, Chengdu, in 2001.

From 2003 to 2004, he was a Research Associate with the Department of
Electrical and Electronic Engineering, The Hong Kong University of Science and Technology. From 2007 to 2008, he was a Visiting Professor with the University of Ulm, Germany. Since 2001, he has been with the School of Information Science and Technology, Southwest Jiaotong University, where he is currently a Professor. His research interests include coding theory, network security, distributed storage, and information processing for big data.

Dr. Tang was a recipient of the National Excellent Doctoral Dissertation Award, China, in 2003, the Humboldt Research Fellowship, Germany, in 2007, and the Outstanding Young Scientist Award by NSFC, China, in 2013. He served as an Associate Editors for several journals, including IEEE Transactions on Information Theory and \textit{IEICE Transactions on
Fundamentals}, and served for a number of technical program committees
of conferences.
\end{IEEEbiography}

\end{document}